\documentclass[12pt,a4paper]{article}
\usepackage{amsmath,caption,bbm,longtable,paralist,amsfonts}
\usepackage[top=1.5in, bottom=1.5in, left=1.1in, right=1.1in]{geometry}

\DeclareCaptionStyle{italic}[justification=centering]{labelfont={bf},textfont={it},labelsep=colon}
\usepackage{graphicx,psfrag,epsf}
\usepackage{enumerate}
\usepackage{natbib}
\usepackage{url} 
\usepackage{bbm, dsfont}
\usepackage{comment}
\usepackage{float}
\usepackage[dvipsnames]{xcolor}
\usepackage{booktabs, subfig, bm, paralist,mathpazo,tikz,todonotes,longtable,microtype,dsfont,rotating} 
\usepackage[pdftex,colorlinks=true]{hyperref}
\definecolor{darkblue}{rgb}{0,0,.6}
\hypersetup{citecolor=darkblue,linkcolor=darkblue,urlcolor=darkblue}

\newcommand{\blind}{0}

\newcommand{\X}{\mathcal{X}}
\newcommand{\Y}{\mathcal{Y}}
\newcommand{\Z}{\mathcal{Z}}

\graphicspath{{plots/}}
\DeclareMathOperator*{\argmin}{\arg\!\min}
\newsavebox\CBox

\definecolor{a0}{rgb}{0.0, 0.5, 0.0}
\definecolor{bistre}{rgb}{0.24, 0.17, 0.12}
\definecolor{amethyst}{rgb}{0.6, 0.4, 0.8}
\definecolor{blue-violet}{rgb}{0.54, 0.17, 0.89}
\definecolor{Rcolor}{RGB}{150,160,190}
\definecolor{blush}{rgb}{0.87, 0.36, 0.51}
\definecolor{brightturquoise}{rgb}{0.03, 0.91, 0.87}
\definecolor{burntorange}{rgb}{0.8, 0.33, 0.0}

\usepackage{orcidlink}
\date{\today}
\AtBeginDocument{}
\usepackage{cleveref}

\begin{document}

\def\spacingset#1{\renewcommand{\baselinestretch}
{#1}\small\normalsize} \spacingset{1}

\if0\blind
{
  \title{\bf Forecasting high-dimensional functional time series: Application to sub-national age-specific mortality}
  \author{Cristian F. Jim\'enez-Var\'on \orcidlink{0000-0001-7471-3845} \ and \ Ying Sun \orcidlink{0000-0001-6703-4270}  \thanks{Postal address: CEMSE Division, Statistics Program, King Abdullah University of Science and Technology, Thuwal 23955-6900, Saudi Arabia. E-mail: ying.sun@kaust.edu.sa} \hspace{.2cm}\\  
  CEMSE Division\\ 
  King Abdullah University of Science and Technology \\ 
  \\
  Han Lin Shang \orcidlink{0000-0003-1769-6430} \\
    Department of Actuarial Studies and Business Analytics \\
    Macquarie University
}
  \maketitle
} \fi


\if1\blind
{
\title{\bf Forecasting high-dimensional functional time series: Application to sub-national age-specific mortality}
} \fi

\bigskip

\begin{abstract}

We study the modeling and forecasting of high-dimensional functional time series (HDFTS), which can be cross-sectionally correlated and temporally dependent. We introduce a decomposition of the HDFTS into two distinct components: a deterministic component and a residual component that varies over time. The decomposition is derived through the estimation of two-way functional analysis of variance. A functional time series forecasting method, based on functional principal component analysis, is implemented to produce forecasts for the residual component. By combining the forecasts of the residual component with the deterministic component, we obtain forecast curves for multiple populations. We apply the model to age- and sex-specific mortality rates in the United States, France, and Japan, in which there are 51 states, 95 departments, and 47 prefectures, respectively. The proposed method is capable of delivering more accurate point and interval forecasts in forecasting multi-population mortality than several benchmark methods considered.

\vspace{.1in}
\end{abstract}

\newpage
\spacingset{1.6}

\section{Introduction}\label{sec:intro}

In recent years, most countries worldwide have seen continuous drops in mortality rates, which are also associated with aging populations. For planning purposes, policymakers from insurance firms and government departments demand more precise mortality forecasts. Several statistical methods have been presented for forecasting age-specific central mortality rates, life-table death counts, or survival functions \citep[see, e.g.,][]{Booth2006, Currie2004, booth2008, Shang2011, BCB23}. 

One of the most outstanding contributions in this field is that of \cite{Lee1992}, who use a principal component method to derive a single time-varying index of the level of mortality rates, from which forecasts are obtained using a random walk with drift. Subsequently, several approaches have modified and extended the Lee-Carter method. For instance, \cite{Renshaw2003} propose the age-period-cohort Lee Carter method; \cite{Hyndman2007} propose a functional data approach along with nonparametric smoothing and high-order principal components; \cite{Girosi2008} and \cite{Wisniowski2015} consider Bayesian techniques for the estimation and forecasting of the Lee-Carter model; and \cite{Li2013} extend the Lee-Carter method to approximate age pattern rotation for long-term projections.

One major drawback of the Lee-Carter method and previous contributions is that they mainly focus on forecasting mortality for a single population. Each population can be further categorized based on gender, state, ethnic group, socioeconomic status, and other factors. Individual forecasts, even when based on identical extrapolating processes, may, in the long run, imply increased divergence in mortality rates, contrary to the expected and observed trend toward global convergence \citep[see, e.g.,][]{Li2012, Pampel2005, HBY13}. 

Joint modeling mortality for two or more populations simultaneously is critical; it considers data correlation and may discriminate between long-term and short-term impacts in mortality evolution. Finally, joint modeling incorporates additional information from other populations that can be used to enhance forecast accuracy. Various proposals have tackled the problem of combining several populations for forecasting. For instance, \cite{Shang2016} proposes multivariate and  multilevel functional data approaches for forecasting age-specific mortality rates for two or more populations in developed countries with high-quality vital registration systems. \cite{Shang2017} and \cite{SH17} advocate employing a grouped functional time series methodology in conjunction with a bootstrap method to provide point and interval forecasts of mortality rates that are correctly aggregated across different disaggregation parameters. 

Functional analysis of variance (ANOVA) is a common option for joint modeling with functional data \citep[][Chapter 13]{RS06}. Functional ANOVA models evaluate the functional impacts of categorical factors by determining how functions differ at different levels of these factors. Functional ANOVA models have proven usefulness in analyzing data in a wide range of applications, such as human tactile perception \citep{Spitzner2003}, menstrual cycle data \citep{Brumback1998}, and circadian rhythms with random effects and smoothing spline ANOVA decomposition \citep{Wang2003}. In particular, \cite{Kaufman2010} establishes a Bayesian framework for functional ANOVA modeling to estimate the effect of geographic regions on Canadian temperatures. 
 
\cite{Sun2012} propose a functional median polish (FMP-ANOVA) modeling as an extension of the univariate median polish proposed by \cite{tukey1977} and \cite{MT77}. FMP-ANOVA computes the functional grand effect and functional main factor effects in an additive model without factor interaction. It is a robust statistical technique for studying the effects of factors on response since it replaces the mean with the median \citep{EH83}. \cite{Sun2012} present a functional rank test to determine the significance of functional main factor effects. Additionally, they prove the robustness of FMP-ANOVA by comparing its performance to the functional ANOVA model fitted by means (FM-ANOVA). 
 
In the functional data analysis (FDA) approach \citep{RS06}, it is assumed that the mortality rate in each year follows an underlying smooth function of age. When mortality rates are collected over time, we refer to the data as functional time series (FTS). Because of observational noise, observed mortality rates are not smooth across ages. We employ a penalized regression spline smoothing with monotonic constraint \citep{Wood1994} to create smooth functions and deal with possible missing data. It considers the shape of the log mortality curve \citep[see also][]{Hyndman2007,Shang2017}. The smooth shape of age-specific mortality rates over age in each year is a distinguishing feature, which can improve short-term forecast accuracy \citep[see,e.g.,][]{BC19, YSR23}.

Smoothing techniques can better capture the underlying trend of mortality changes, reducing the impact of missing values and measurement noise in various sub-national series \citep{YSR23}. Functional ANOVA models provide an insightful decomposition of functional mortality rates into a deterministic component (such as populations or states as in the functional factor effects) and a functional residual component that varies over time. The functional residuals are inputs for mortality forecasting. 
 
Hence, functional time series forecasting methods often require efficient data reduction algorithms, since functional residuals are infinite-dimensional functions. For example, the most often used approach for this purpose is functional principal component analysis (FPCA). FPCA represents functional data on their eigenfunction basis. Several studies discuss FPCA, in particular, \cite{Hall2006} and \cite{Hall20062} for theoretical properties, \cite{Viviani2005} and \cite{Locantore1999} for empirical applications,  \cite{Shang14} and \cite{WCM16} for surveys. 

In the FTS literature, different approaches are presented regarding the construction of prediction intervals. For instance, \cite{Antoniadis2006, Antoniadis2016} execute one-step-ahead prediction using a nonparametric wavelet kernel; pointwise prediction intervals are produced using a re-sampling approach. Some other contributions, such as \cite{Rana2016} and \cite{Vilar2018}, use model-based bootstrap procedures for constructing pointwise prediction intervals for one-step-ahead prediction. Such approaches are mainly based on assumptions on the data-generating process under the functional autoregressive model of order 1 (FAR(1)).

\cite{Aue2015} introduces an approach for constructing prediction intervals, in which a tuning parameter is selected to achieve the smallest distance between the empirical and nominal coverage probabilities based on the in-sample data. \cite{Paparoditis2018} presents a sieve bootstrap approach for FTS that employs the vector autoregressive (VAR) representation of the time series of Fourier coefficients appearing in the Karhunen-Lo\`{e}ve expansion of the functional process. For a stationary series, the VAR representation can be written forward and backward. \cite{Paparoditis2021} introduce a sieve-bootstrap approach for constructing prediction bands for FTS that considers the different sources of error, including the model misspecification error, affecting the conditional distribution of the estimated functional residuals.

In the context of high-dimensional functional time series (HDFTS), \cite{ZD23} derived Gaussian and multiplier bootstrap approximations for sums of HDFTS. By utilizing these approximations, they were able to construct joint simultaneous confidence bands for the mean functions and develop a hypothesis test to assess whether the mean functions in the panel dimension exhibit parallel behavior.  \cite{HNT231} investigated the representation of HDFTS using a factor model, identifying conditions on the eigenvalues of the covariance operator crucial for establishing the existence and uniqueness of the factor model. \cite{GSY19} adopted a two-stage approach combining truncated principal component analysis and a separate scalar factor model for the resulting panels of scores, while \cite{HNT232} introduced a functional factor model with a functional factor loading and a vector of real-valued factors, and \cite{GQW22} considered a functional factor model with a real-valued factor loading and a functional factor. Additionally, \cite{TSY22} studied clustering problems for age-specific mortality rates, an example of HDFTS, and \cite{LLS23} proposed hypothesis tests for change point detection, change point estimation, and clustering of change points in HDFTS using an information criterion.

In the realm of estimating HDFTS models, \cite{GQ23} introduced a three-step procedure that incorporates a novel functional stability measure, the non-asymptotic properties of functional principal component analysis (FPCA), and a regularization approach for estimating autoregressive coefficients. Moreover, \cite{TLG+24} proposed dynamic weak separability as a characterization of the two-way dependence structure in multivariate functional time series, developing a unified framework for functional graphical models and dynamic principal component analysis. Finally, \cite{CCQ+23} presented a three-step framework for statistical learning of HDFTS with errors, incorporating autocovariance-based dimension reduction and a novel block regularized minimum distance estimation. 
 
 In this paper, we propose an innovative forecasting approach for HDFTS. Our method begins with the estimation of a two-way functional ANOVA model, that decomposes the HDFTS into a deterministic component and functional residuals that vary over time. Such estimation of the functional ANOVA model can be carried out by either means or medians. In particular, in the presence of functional outliers, FMP-ANOVA has been proven to be robust against such outliers \citep{Sun2012}. After removing the deterministic component, a functional principal component regression is deployed to model and forecast the functional residual component. Finally, we obtain forecast curves for several populations by combining the forecasts of the functional residuals that vary over time with the deterministic components. The proposed FTS forecasting method based on a decomposition from the estimation of a two-way analysis of variance model is compared with existing methods such as the factor models from \cite{GSY19} and \cite{HNT232}. 

We investigate the proposed method's point and interval forecast accuracies using age- and sex-specific mortality rates from the US, France, and Japan. For the US, we consider 51 states from 1959 to 2020; for France, 95 departments from 1968 to 2021; and for Japan, 47 prefectures from 1975 to 2020. We consider the mean relative absolute prediction error (MAPE) and root mean relative squared prediction error (RMSPE) for evaluating point forecast accuracy. For comparing interval forecast accuracy, we study the difference between the empirical and nominal coverage probabilities and the mean interval score of \cite{Gneiting2007} and \cite{GK14}.

The remainder of this paper is structured as follows. In \S~\ref{sec:empirical}, we present the US, French, and Japanese sub-national mortality rates. In \S~\ref{sec:methods}, we introduce a FTS forecasting method for producing point and interval forecasts. The proposed FTS forecasting method is based on both functional ANOVA decomposition and FPCA.  We evaluate and compare point and interval forecast accuracies with existing joint time series forecasting methods in \S~\ref{PF_comp} and \S~\ref{IF_comp}, respectively.  \S~\ref{sec:conclusion} concludes and offers some ideas on how the methodology presented can be extended.

\section{Age-specific mortality rates in the United States, France, and Japan}\label{sec:empirical}

\begin{figure}[!htbp]
\centering
\includegraphics[width=8.0 cm]{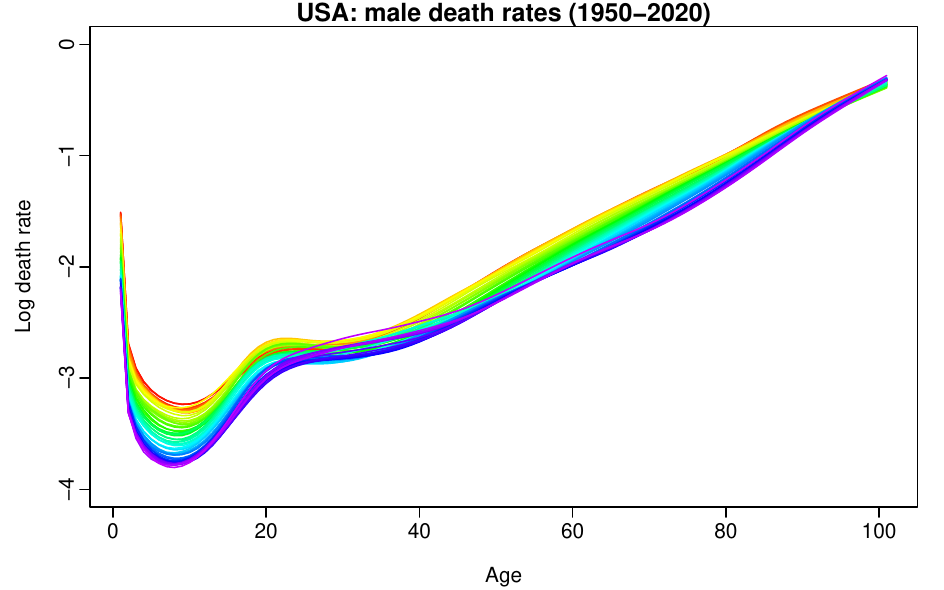}
\qquad
\includegraphics[width=8.0 cm]{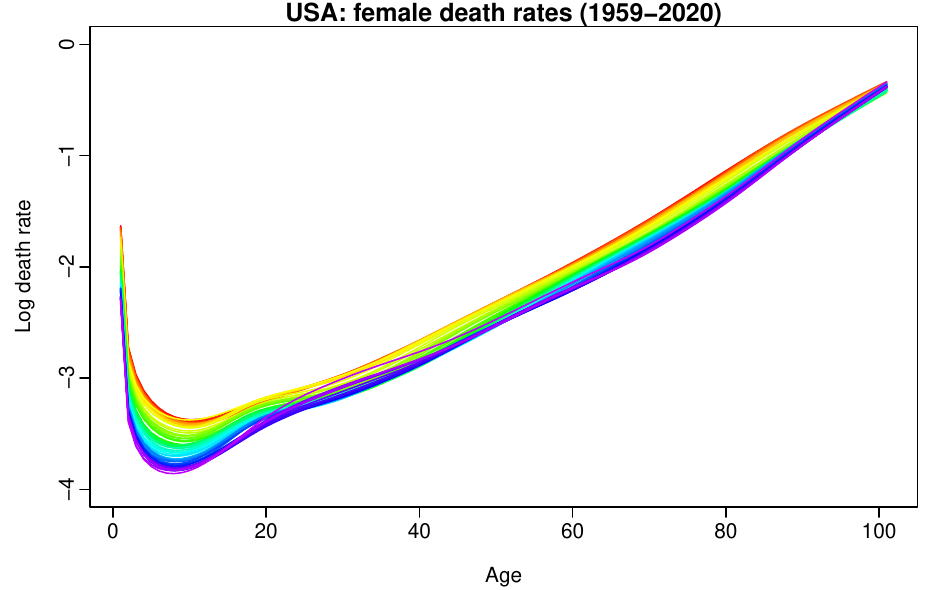}
\\
\includegraphics[width=8.0 cm]{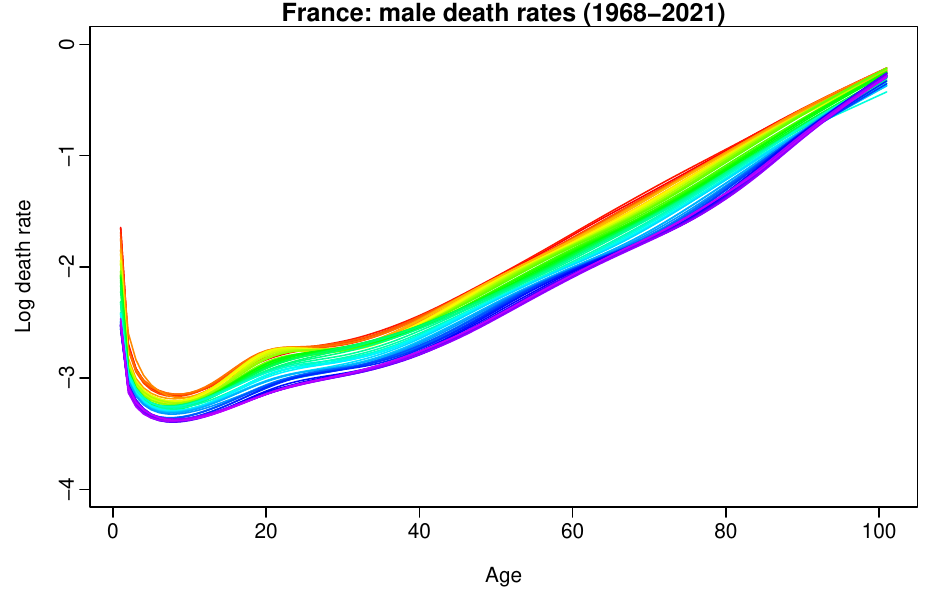}
\qquad
\includegraphics[width=8.0 cm]{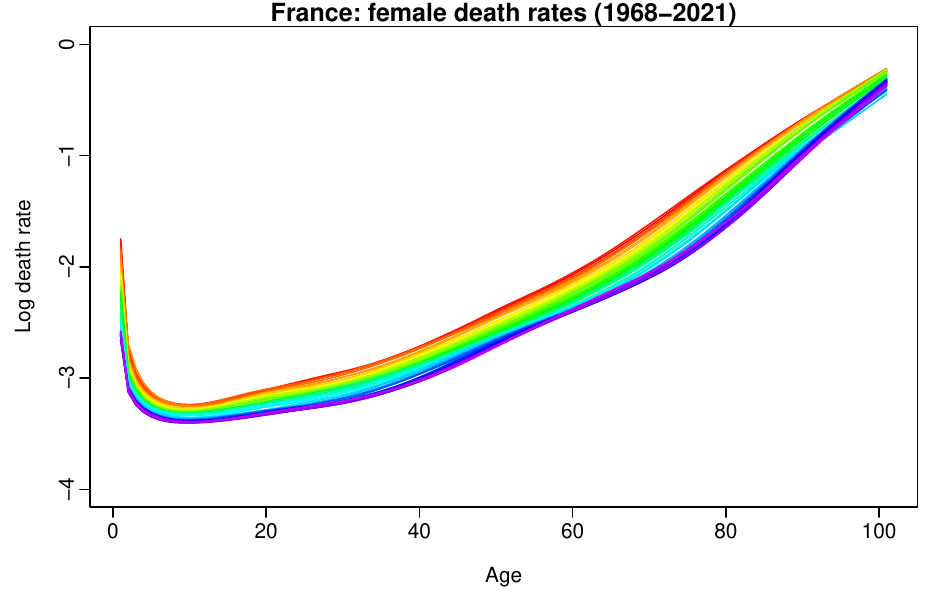}
\\
\includegraphics[width=8.0 cm]{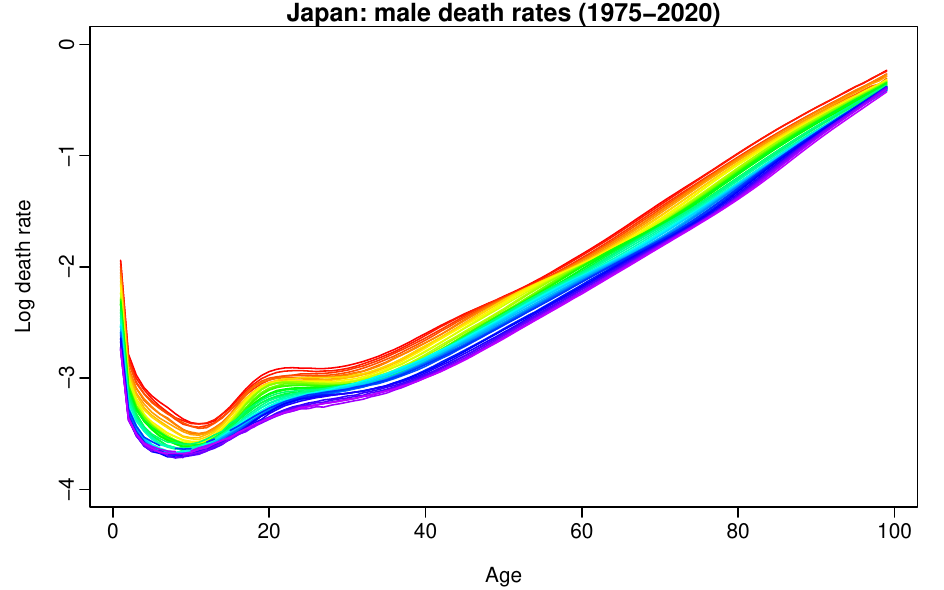}
\qquad
\includegraphics[width=8.0 cm]{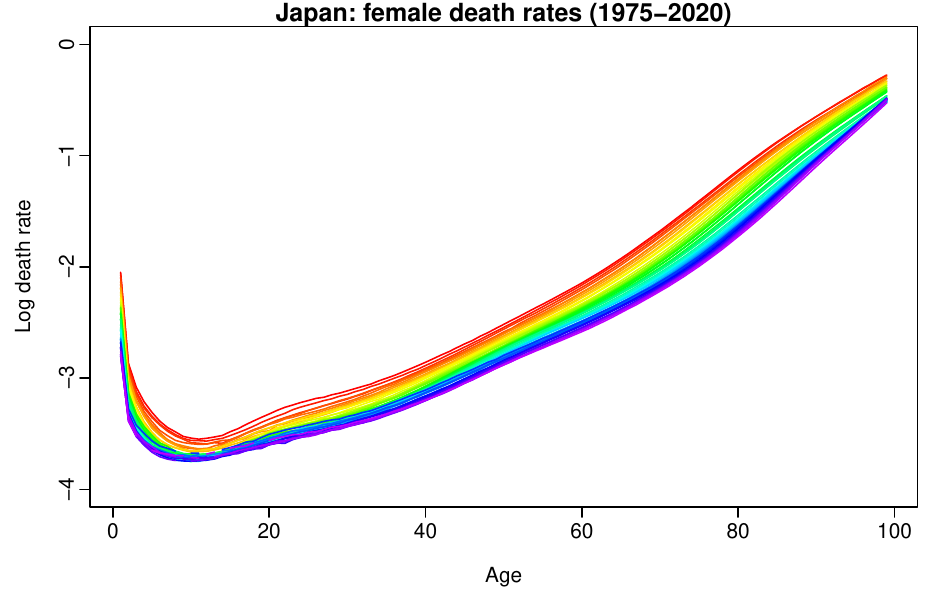}
\caption{\small{The smoothed age-specific $\log_{10}$ mortality rates between 1959 and 2020 in the US, between 1968 and 2021 in France, and between 1975 and 2020 in Japan. Curves are ordered chronologically using the rainbow color palette, the oldest curves are shown in red, with the most recent curves in violet.}}\label{fig:1}
\end{figure}

The United States Mortality Database documents a historical set of complete state-level life tables designed to foster research on geographic variations in mortality across the US and to monitor trends in health inequalities \citep{USMD23}. This data set currently includes complete and abridged life tables by sex for each of the nine US  Census Divisions, four Census Regions, 50 States, and the District of Columbia, for each year between 1959 and 2020, with mortality up to age 110. To motivate the discussion, consider the first row of Figure~\ref{fig:1} showing annual age- and sex-specific $\log_{10}$ mortality rates for the US.

The French Human Mortality Database \citep{Bonnet20} allows the general public to assess mortality data by region. We are interested in the dynamic changes in the mortality rates in France at the departmental level and, particularly, in the age-specific mortality rates in a single-year interval by sex. In Europe, France has 95 departments, with mortality data up to age 110. We examine age groupings ranging from 0 to 100 in single years of age, with the last age group including all ages above 100. These departments are Seine and Seine et Oise, which have been removed from our further analyses.  We examine age groupings ranging from 0 to 100 in single years of age, with the last age group including ages above 100.  The annual age-specific $\log_{10}$ mortality rates for the French females and males between 1968 and 2021 are shown in the second row of Figure~\ref{fig:1}.

We investigate the Japanese age-specific mortality rates from 1975 to 2020, as obtained from the \cite{JMD23}. The mortality rates are the ratios of death counts to population exposure in the relevant year for the given age (based on a one-year age group). We examine age groupings ranging from 0 to 98 in single years of age, with the last age group including all ages at and above 99. We do not have the same truncation age as for the previous cases due to missing values in some of the prefectures for older ages.  The annual age-specific $\log_{10}$ mortality rates for Japanese females and males between 1975 and 2020 are shown in the last row of Figure~\ref{fig:1}.  Due to ease of interpretability, we refer by $\log$ to the base $10$ logarithm ($\log_{10}$).

\section{Methodology}\label{sec:methods}

We introduce two decompositions based on functional analysis of variance in \S~\ref{FMP}, the functional median polish (FMP-ANOVA) approach of \cite{Sun2012}, and the functional mean analysis of variance (FM-ANOVA). \S~\ref{FM} reports the forecasting method for FTS based on FPCA. \S~\ref{PF_FMP} describes the proposed FTS forecasting method based on the estimation of a functional analysis of variance decomposition and FPCA. \S~\ref{intervals} explains a sieve bootstrap methodology to obtain prediction intervals for mortality curves. For consistency, we refer to departments and prefectures as states.

\subsection{A two-way functional ANOVA decomposition} \label{FMP}

Let $\Y_{t,s}^g(u)$ be the $\log_{10}$ mortality rate for age $u$, state $s$, gender $g$ at year $t$. By treating age $u$ as a continuum, a two-way functional ANOVA model can be estimated for FTS. In this way,  $\Y_{t,s}^{g}(u)$ can be decomposed as: 
\begin{equation}
   \Y_{t,s}^{g}(u)=\mu(u)+\alpha_s(u)+\beta^g(u)+\X^{g}_{t,s}(u), \qquad ,
   \label{FMP/FM}
\end{equation}
where, $\mu(u)$ denotes the functional grand effect, $\alpha_{s}(u)$ denotes the functional row effect and, $\beta^{g}(u)$ denotes the functional column effect.  Although $u$ ($u\in \mathcal{I}\subset \mathds{R}$) is a continuous variable, it can only be observed at a set of grid points, such as $(u_1,\ldots, u_p)$. For each state $s$ and gender $g$, we considered replicates the years $t$ with time horizon, for $t=1,\ldots, T$, $s=1,\ldots,n_s$, and $g=\text{Female}(F),\text{Male}(M)$. Finally, $\bm{\X}_{s}^{g}(u)=[\X_{1,s}^{g}(u),\dots,\X_{T,s}^{g}(u)]$ denotes the functional residual process for the state $s$ and gender $g$.  The model described in~\eqref{FMP/FM}, can be estimated by two approaches: The FM-ANOVA model and the FMP-ANOVA approach of \cite{Sun2012}. In particular, if the model in ~\eqref{FMP/FM}, is estimated through the FMP-ANOVA decomposition, it satisfies that  $\forall u \in \mathcal{I}$, $\text{median}_s\{ \alpha_s(u)\}=0$, $\text{median}_g\{ \beta^{g}(u)\}=0$, $\text{median}_s\{ \X^{g}_{t,s}(u)\}=\text{median}_g\{ \X^{g}_{t,s}(u)\}=0$ for all $t$ \citep{Sun2012}.

Alternatively, the functional ANOVA model in \eqref{FMP/FM} can be fitted by means (FM-ANOVA) with \citep{RS06,Sun2012} 
\begin{align*}
    \widehat{\mu}(u)&=\frac{1}{T}\sum_{s=1}^{n_s}\sum_{g=1}^{n_g}\sum_{t=1}^{T} \Y_{t,s}^{g}(u)\\
                   \widehat{\alpha}_s(u)&= \frac{1}{T}\sum_{g=1}^{n_g}\sum_{t=1}^{T} \Y_{t,s}^{g}(u)-\widehat{\mu}(u)\\
                    \widehat{\beta}^{g}(u)&= \frac{1}{T}\sum_{s=1}^{n_s}\sum_{t=1}^{T} \Y_{t,s}^{g}(u)-\widehat{\mu}(u).
\end{align*}

In the FM-ANOVA decomposition, there exist some identifiability constraints, so that for all $u \in \mathcal{I}$, $\sum_{s=1}^{n_s}\alpha_s(u)=\sum_{g=1}^{n_g}\beta^{g}(u)=0$, and $\sum_{s=1}^{n_s} \X^{g}_{t,s}(u)=\sum_{g=1}^{n_g}\X^{g}_{t,s}(u)=0$ for all $t$ \citep[][Chapter 13]{RS06}.

\subsection{Functional time series forecasting method}\label{FM}

We introduce the FTS forecasting approach based on the FPCA which relies on an accurate estimate of the covariance function. A brief description of the estimation of the covariance function is given in \S~\ref{cov}. \S~\ref{SFPC} presents the basic ideas on  FPCA designed for FTS forecasting. 

\subsubsection{Covariance function} \label{cov}

For a given state $s$ and gender $g$, denote $\X_{t,s}^g(u)$ a stochastic process defined on a compact set~$\mathcal{I}$, with finite variance $\int_{\mathcal{I}}\mathds{E}\left((\X_{t,s}^{g}(u))^2 \right) < \infty$. Furthermore, $\X_{t,s}^g(u)$ can be seen as a stationary ergodic FTS exhibiting stationarity and ergodicity. In essence, the statistical features of a stochastic process will not vary over time, and they can be obtained from a single, sufficiently long sample of the process. The  covariance function  of $\X_{t,s}^g(u)$ is defined  to be the function $C(u,v):\mathcal{I}\times \mathcal{I} \rightarrow \mathds{R} $, such that
\begin{equation}
    \begin{aligned}
    C(u,v)&=\text{Cov}\left(\X_{t,s}^g(u),\X_{t,s}^g(v) \right)\\
    &=\mathds{E}\left[\X_{t,s}^g(u)\X_{t,s}^g(v)\right].
    \label{Cov}
\end{aligned}
\end{equation}

By assuming $\X_{t,s}^g(u)$ is a continuous and square-integrable function, the function $C(u,v)$ induces the kernel operator $L^2(\mathcal{I})\rightarrow L^2(\mathcal{I})$, $\phi \rightarrow C\phi $, given by
\begin{equation*}
    (C\phi)(u)=\int_{\mathcal{I}} C(u,v)\phi(v)dv.
\end{equation*}

\subsubsection{Functional principal component analysis} \label{SFPC}

With the covariance function defined in~\eqref{Cov},  and via Mercer's lemma \citep{Mercer}, there exists an orthonormal sequence $(\phi_k)$ of continuous function in $L^2(\mathcal{I})$ and a non-increasing sequence of positive numbers $\theta_k$, such that 
\begin{equation*}
    C(u,v)=\sum_{k=1}^{\infty} \theta_{k} \phi^{g}_{k}(u)\phi^{g}_{k}(v),
\end{equation*}
where, the orthonormal functions $[\phi^{g}_1(u),\phi^{g}_2(u),\ldots]$ are referred as the functional principal components. We can project the  FTS $\X_{t,s}^g(u)$ onto a collection of orthogonal functional principal components $\phi^{g}_k$ via the inner product in the corresponding Hilbert space. This leads to the Karhunen-Lo\`{e}ve expansion of  the realization of the stochastic process $\X_{t,s}^g(u)$ that can be expressed as,
\begin{equation*}
\X_{t,s}^g(u)=\overline{\X}_s^g(u)+\sum_{k=1}^{\infty} \Gamma_{k,t,s}^g\phi^{g}_{k,s}(u),
\end{equation*}
where $\Gamma^{g}_{k,t,s}=\bigl \langle \X_{t,s}^g(u)-\overline{\X}_s^g(u),\phi^{g}_{k,s}(u) \bigr \rangle$, denotes the $k^{th}$ set of principal component scores for time $t$.

\subsection{Point forecasts based on a two-way functional ANOVA decomposition}\label{PF_FMP}

Define $\Y^{g}_{t,s}(u), t=1, 2, \ldots, T$ as a set of smoothed sub-national mortality rates functions, with $T$ representing the number of total smoothed curves. For instance, assume that each $\Y^{g}_{t,s}(u)$ is a square-integrable function defined at the same interval of age. Through a two-way functional ANOVA decomposition approach, as  described in \S~\ref{FMP}, $\Y^{g}_{t,s}(u)$ can be decomposed into two main components: 
\begin{inparaenum}
   \item[(1)] a deterministic component; and 
   \item[(2)] a residual component that varies over time.
\end{inparaenum}
The estimated deterministic component includes the functional grand effect $\widehat{\mu}(u)$, functional row effect $\widehat{\alpha}_{s}(u)$, and functional column effect $\widehat{\beta}^{g}(u)$. The estimated component that varies over time refers to the functional residuals $\widehat{\bm{\X}}^{g}_{s}(u)=[\widehat{\X}_{1,s}^g,\dots,\widehat{\X}_{T,s}^g]$. That is
\[
    \underbrace{\widehat{\X}^{g}_{t,s}(u)}_{\text{Residual }}=\Y_{t,s}^g(u) - \underbrace{[\widehat{\mu}(u)+\widehat{\alpha}_{s}(u)+\widehat{\beta}^{g}(u)]}_{\text{deterministic}}.
\]

When estimating the two-way functional ANOVA model in equation \eqref{FMP/FM} using either the FM-ANOVA or FMP-ANOVA approaches, it is important to note that these decompositions separate the functional time series (FTS) into a deterministic component and a residual component. However, regardless of the chosen estimation method, these decompositions ensure the exact reconstruction of the original data.

Once the deterministic components are removed from the decomposition of $\Y^{g}_{t,s}(u)$, the residual components are used for FPCA as described in \S~\ref{FM}.  To incorporate the correlation between the residual components, $\bm{\widehat{\X}}^g_{s}(u)$ can be stacked as for a given state $s$ for female and male populations  $\left[ \text{Let}~\bm{\widehat{\X}}_s(u) = [\bm{\widehat{\X}}^{\text{F}}_{s}(u), \bm{\widehat{\X}}^{\text{M}}_{s}(u)]^{\top}\right]$. Let $\widehat{C}(u,v)$ be an estimator of the covariance function defined in \eqref{cov} for the estimated residuals $\widehat{\X}_{t,s}(u)=[\widehat{\X}^{\text{F}}_{t,s}(u),\widehat{\X}^{\text{M}}_{t,s}(u)]^\top$. via Mercer's lemma  $\widehat{C}(u,v)$ can be approximated by,
\begin{equation*}
    \widehat{C}(u,v)=\sum_{k=1}^{T}\widehat{\theta}_{k} \widehat{\phi}_{k}(u)\widehat{\phi}_{k}(v).
\end{equation*}
Through functional principal component analysis, it can be decomposed as follows:
\begin{align*}
    \widehat{\X}_{t,s}(u) &= \overline{\widehat{\X}}_s(u) + \sum^{\infty}_{k=1}\widehat{\Gamma}_{k,t,s}\widehat{\phi}_{k,s}(u)\\
    &= \overline{\widehat{\X}}_s(u) + \sum^{K}_{k=1}\widehat{\Gamma}_{k,t,s}\widehat{\phi}_{k,s}(u) + \widehat{\varepsilon}_{t,s}(u),
\end{align*}
where $[\widehat{\phi}_{1,s}(u),\ldots,\widehat{\phi}_{K,s}(u)]$ is a set of orthogonal basis functions commonly known as a functional principal component for the $s$\textsuperscript{th} state,  with $\widehat{\Gamma}_{k,t,s}$ as their related principal component scores for $k=1,\ldots, K,t=1,\dots, T$; and $\widehat{\varepsilon}_{t,s}(u)$ denotes the model truncation error function with mean zero and finite variance. We select $K$ by the eigenvalue ratio (EVR) criterion of  \citep{LRS21}, such estimator is obtained simply by minimizing the ratio of two adjacent eigenvalues arranged in descending order.  

\begin{equation*}
  K =  \argmin_{k: 1 \leq k \leq k_{\max}} \Biggl\{ \frac{\lambda_{k+1}}{\lambda_k} \times \mathbbm{1}\{ \lambda_k > \tau \} + \mathbbm{1}\{ \lambda_k < \tau\} \left. \Biggr\} \right.,
\end{equation*}
where,$\mathbbm{1}\{\cdot\}$ represents the binary indicator function. Customary, $\tau=0.001$ and $k_{\max}$ can be set as~$T$. The selection of $K$ is a crucial step, as the accuracy of the forecasting method relies heavily on choosing the appropriate number of principal components. It is important to note that \cite{LY12} (Theorem 2) has demonstrated that the selected value of $K$ is consistent and optimal.

There are different alternatives for determining the number of retained components:
\begin{inparaenum}
\item[(1)] scree plots or the fraction of variance explained by the first few functional principal components \citep{Chiou12};
\item[(2)] pseudo-versions of Akaike information criterion and Bayesian information criterion \citep{YMW05};
\item[(3)] predictive cross validation leaving out one or more curves \citep{RS06};
\item[(4)] bootstrap methods \citep{HV06}; 
\item[(5)] eigenvalue ratio criterion \citep{AH13}.
\end{inparaenum}

Collectively modeling multiple populations requires truncating the $K$\textsuperscript{th} functional principal components of all time series 
\[
\bm{\widehat{\X}}_{t,s}(u) \approx \bm{\widehat{\Phi}}_s(u)\bm{\widehat{\Gamma}}_{t,s},
\]
where $\bm{\widehat{\X}}_{t,s}(u) = [\widehat{\X}_{t,s}^{\text{F}}(u),\widehat{\X}_{t,s}^{\text{M}}(u)]^{\top}$, and

\begin{equation*}
\bm{\widehat{\Gamma}}_{t,s}=\left[\widehat{\Gamma}^{\text{F}}_{1,t,s}, \ldots, \widehat{\Gamma}^{\text{F}}_{K,t,s},\widehat{\Gamma}^{\text{M}}_{1,t,s},\ldots, \widehat{\Gamma}^{\text{M}}_{K,t,s}\right]^{\top},
\end{equation*}
\noindent is a $((2 \times K) \times 1)$ vector of principal component scores.  $\bm{\widehat{\Phi}}_s(u)$ is a $2 \times (2 \times K)$ matrix that contains the associated basis functions, in which $\bm{\widehat{\Phi}}_s(u)$ is given by
\begin{equation*}
    \bm{\widehat{\Phi}}_s(u)=\begin{pmatrix}
\widehat{\phi}^{\text{F}}_{1,1}(u) & \ldots &\widehat{\phi}^{\text{F}}_{K,1}(u)  &  & \ldots &  \\ 
 & \ldots &  & \widehat{\phi}^{\text{M}}_{1,2}(u) &\ldots  & \widehat{\phi}^{\text{M}}_{K,2}(u)  
 \end{pmatrix}.
\end{equation*}
By conditioning on $\bm{\widehat{\Phi}}_s(u)$, we can now obtain the $h$-step-ahead point forecasts as follows
\begin{align*}
\bm{\widehat{\X}}_{T+h|T,s}(u)&=\mathds{E}\Big[\bm{\widehat{\X}}_{T+h,s}(u)\big|\bm{\widehat{\X}}_{1,s}(u),\ldots,\bm{\widehat{\X}}_{T,s}(u);\bm{\widehat{\Phi}}_s(u)\Big]\\
&=\bm{\overline{\widehat{\X}}}_s(u)+\bm{\widehat{\Phi}}_s(u)\widehat{\bm{\Gamma}}_{T+h|T,s},
\end{align*}
where the empirical mean function $\bm{\overline{\widehat{\X}}}_s(u)=[\overline{\widehat{\X}}_s^{\text{F}}(u), \overline{\widehat{\X}}_s^{\text{M}}(u)]$. In this paper, we use the univariate time series forecasting method of \cite{Hyndman2009} to obtain the forecast principal component score $\widehat{\bm{\Gamma}}_{T+h|T,s}$ \citep[see also][]{Shang2017, Shang2021}.  Alternative to a univariate time series method, a multivariate time series forecasting method, such as VAR, can also be implemented in this step. In Appendix~\ref{A1}, we compare the univariate and multivariate time series forecasting approaches for the three datasets presented in~\S~\ref{PF_comp} when computing point forecast. 

Once the forecasted functional residuals are obtained, we add back the deterministic component from the functional ANOVA decomposition used. As this does not vary over time,  the overall $h$-step-ahead point forecast is defined as
\[
    \widehat{\Y}_{T+h|T,s}^{g}(u)=\widehat{\mu}(u)+\widehat{\alpha}_{s}(u)+\widehat{\beta}^{g}(u) + \widehat{\X}_{T+h|T, s}^g(u).
\]

\subsection{Construction of prediction intervals} \label{intervals}

In the functional-ANOVA decompositions described in \S~\ref{FMP},  we consider joint modeling for both female and male populations to obtain the functional residuals. In this section, we use the functional residuals for each population separately to compute prediction intervals for quantifying forecast uncertainty using the approaches proposed by \cite{Paparoditis2021} and \cite{Aue2015}.  The final prediction intervals are generated after adding back the deterministic components removed from the functional ANOVA decomposition. The procedure can be described as follows.

\begin{enumerate}
\item[1)] \label{step1} Center the observed functional time series by calculating $\Z_{t,s}^g(u) = \widehat{\X}_{t,s}^g(u) - \overline{\widehat{\X}}_{s}^g(u)$, where $\overline{\widehat{\X}}_s^g(u) = \frac{1}{T}\sum^T_{t=1}\widehat{\X}_{t,s}^g(u)$.

\item[2)] Apply the FPCA decomposition to $\bm{\Z}_s^g(u)=[\Z_{1,s}^g(u),\dots,\Z_{T,s}^g(u)]$ to obtain a set of estimated functional principal components and their associated scores. 

\item[3)] Fit a vector autoregression of order $p$, VAR($p$), process to the ``forward" series of the estimated scores; that is,
\[
\widehat{\Gamma}_{m,s}^g = \sum^p_{j=1}A_{j,p}\widehat{\Gamma}_{m-j,s}^g + \epsilon_{m,s}^g, \qquad m=p+1,\dots,T,
\]
with $\epsilon_{m,s}^g$ being the residuals and $A_{j,p}$ denotes the forward VAR($p$) coefficient. Generate 
\[
\widehat{\Gamma}_{T+h,s}^{g,*} = \sum^p_{j=1}A_{j,p}\widehat{\Gamma}_{T+h-j,s}^{g,*} + \epsilon_{T+h,s}^{g,*},
\]
where we set $\widehat{\Gamma}_{T+h-j}^{g,*}=\widehat{\Gamma}_{T+h-j}$ if $T+h-j\leq T$ and $\epsilon_{T+h,s}^{g,*}$ is independent and identically distributed (iid) resampled from the set of centered residuals $(\epsilon_{m,s}^g - \overline{\epsilon}_{s}^g)$, $\overline{\epsilon}_{s}^g = (T-p)^{-1}\sum^T_{m=p+1}\epsilon_{t,s}^g$. Compute
\[
\widehat{\X}_{T+h,s}^{g,*}(u) = \overline{\widehat{\X}}_s^g(u) + \sum^K_{k=1}\widehat{\Gamma}_{k,T+h,s}^{g,*}\widehat{\phi}_{k,s}^g(u)+U_{T+h,s}^{g,*}(u),
\]
where $U_{T+h,s}^{g,*}(u)$ is iid resampled from the set $\{U_{t,s}^{g}(u) - \overline{U}_{s}^{g}(u), t=1,2,\dots,T\}$, $\overline{U}_s^g(u) = T^{-1}\sum^T_{t=1}U_{t,s}^g(u)$ and $U_{t,s}^{g}(u) = \widehat{\X}_{t,s}^g(u) - \sum^K_{k=1}\widehat{\Gamma}_{k,t,s}^g \widehat{\phi}_{k,s}^g(u)$.   In this step,  it is assumed that the basis functions $\widehat{\phi}_{k,s}^g(u)$ remain fixed during the bootstrapping process, which is conducted using the forward VAR of the principal component scores.
    
\item[4)] \label{BS} Fit a VAR($p$) process to the ``backward" series of the estimated scores; that is,
\[
\widehat{\Gamma}_{\nu,s}^g = \sum^p_{j=1}B_{j,p}\widehat{\Gamma}_{\nu+j,s}^g+\xi_{\nu,s}^g \qquad, \nu = 1, 2, \dots, T-p,
\]
where $B_{j,p}$ denotes the backward VAR($p$) coefficient.

\item[5)] Generate a pseudo-time series of the scores $\{\widehat{\Gamma}_{1,s}^{g,*},\dots, \widehat{\Gamma}_{T,s}^{g,*}\}$ by setting $\widehat{\Gamma}_{t,s}^{g,*} = \widehat{\Gamma}_{t,s}^{g}$ for $t=T, T-1,\dots, T-w+1$, and by using for $t=T-w, T-w-1, \dots, 1$, the backward VAR representation $\widehat{\Gamma}_{\nu,s}^{g,*} = \sum^p_{j=1}B_{j,p}\widehat{\Gamma}_{\nu+j,s}^{g,*}+\xi_{\nu,s}^{g,*}$.

\item[6)] Generate a pseudo-functional time series $\{\widehat{\X}_{1,s}^{g,*},\dots,\widehat{\X}_{T,s}^{g,*}\}$ as follows. For $t=T, T-1,\dots, T-w+1$ set
\[
\widehat{\X}_{t,s}^{g,*}(u) = \overline{\widehat{\X}}_s^{g}(u) + \sum^K_{k=1}\widehat{\Gamma}_{k,t,s}^{g}\widehat{\phi}_{k,s}^g(u)+U_{t,s}^g(u),
\]
and $w$ is a user-specific tuning parameter, while for $t=T-w, T-w-1,\dots,1$, use the obtained backward pseudo-scores $\widehat{\Gamma}_{1,s}^{g,*},\dots, \widehat{\Gamma}_{T-w,s}^{g,*}$ and calculate
\[
\widehat{\X}_{t,s}^{g,*}(u) = \overline{\widehat{\X}}_s^{g}(u) + \sum^K_{k=1}\widehat{\Gamma}_{k,t,s}^{g, *}\widehat{\phi}_{k,s}^g(u)+U_{t,s}^{g,*}(u).
\]
where $U_{t,s}^{g,*}(u)$ are iid pseudo-elements. In \cite{Paparoditis2021}, $w=1$, that is, the bootstrap samples are the same as the most recent curve.

\item[7)] For each bootstrapped $\widehat{\X}_{t,s}^{g,*}(u)$, we apply a functional time series forecasting method to obtain its $h$-step-ahead forecast, denoted by $\widehat{\X}_{T+h|T,s}^{g,*}(u)$. 

\item[8)] The model calibration error, $\omega_{T+h,s}^{g,*}(u) = \widehat{\X}_{T+h,s}^{g,*}(u)-\widehat{\X}_{T+h|T,s}^{g,*}(u)$, is the difference between the VAR extrapolated forecasts in Step 3) and the model-based forecasts in Step 7). 

\item[9)] We compute the pointwise standard deviation based on $(\omega_{T+h,s}^{g,1},\dots, \omega_{T+h,s}^{g,B})$ where $B$ denotes the total number of bootstrap samples. As in \cite{Aue2015}, we search for an optimal tuning parameter $\delta$, where the symmetric prediction interval $(-\delta \times \text{sd}[\omega_{T+h,s}^{g,1},\dots, \omega_{T+h,s}^{g,B}], \delta \times \text{sd}[\omega_{T+h,s}^{g,1},\dots, \omega_{T+h,s}^{g,B}])$ achieves the smallest coverage probability difference between the empirical and nominal coverage probabilities based on the in-sample data.

\item[10)] Using the same functional time-series forecasting method, we apply it to the original functional time series to obtain the $h$-step-ahead forecast, denoted by $\widehat{\X}_{T+h|T,s}^g(u)$. The symmetric prediction interval can be obtained from Step 9), with the selected $\delta$.

\item[11)]  We add the deterministic component from the functional ANOVA decomposition used to obtain the functional residuals you started with in step~\ref{step1} to the bootstrap samples obtained in \S~\ref{BS}. The prediction interval of age-specific mortality curves is
\begin{equation*}
    \widehat{\Y}^{g,\ell}_{T+h|T,s}(u)=\widehat{\mu}(u)+\widehat{\alpha}_{s}(u)+\widehat{\beta}^g(u)+\widehat{\X}^{g, \ell}_{T+h|T,s}(u),
\end{equation*}
where $\ell$ symbolizes either the lower or upper bound.
\end{enumerate}

\section{Forecast accuracy evaluation of sub-national mortality data}

The forecasting method based on a functional ANOVA decomposition and  FPCA is applied to the three datasets, namely the age- and sex-specific mortality rates for the US, France, and Japan. In \S~\ref{PF_comp}, we explain a forecasting scheme for computing point forecasts and evaluating accuracy using two measures of point forecast error. In \S~\ref{IF_comp}, we focus on the interval forecasts and the computation of empirical coverage probability. We present the point and interval forecasting results in \S~\ref{Comp:PF} and \S~\ref{Comp:IF_FMP} respectively.  We evaluate and compare our proposed method based on the estimation of the functional ANOVA model (FMP-ANOVA and FM-ANOVA) with two benchmark functional factor models \citep{GSY19, HNT232}. In the Appendix~\ref{A2}, we compare the proposed method with a na\"{i}ve approach by treating each population independently.

\subsection{Point forecast evaluation} \label{PF_comp}

We consider the rolling window scheme to assess the point forecast as described in \citet[][Chapter 9]{zivot2006}. The procedure is carried out as follows:

\begin{enumerate}
    \item[1)]  The mortality curves are decomposed through a two-way functional ANOVA described in~\S~\ref{FMP} into deterministic and functional residual components. The two factors are the states $s$ and two populations ($F$ and $M$). The functional residual curves $\bm{\widehat{\X}}_s^g(u) = [\widehat{\X}_{1,s}^{g}(u),\dots,\widehat{\X}_{T,s}^g]$ are the ones obtained after removing all deterministic components.
    
    \item[2)]  We start by performing a $h$-step-ahead point forecast of the functional residual component. Then, we add the deterministic components to obtain the point forecast of the future curves.

    \item[3)] To compute each of the $h$-step-ahead point forecasts, for $h=1,\ldots, H$, we proceed as follows: for the $h$-step-ahead point forecast, we consider a rolling window as a training set of size $T$ and produce a $(T+h)$-step-ahead point forecast, and add back the deterministic components.
 
    \item[4)] The process iterates over $h$, and the training set rolls one-step-ahead each time until $T+H$.
\end{enumerate}

 We use the relative root mean squared prediction error (RMSPE) and the mean absolute prediction error (MAPE) to evaluate the point forecast accuracy. They measure how close the forecasts are compared to the actual values of the forecast variable. We compute the relative RMSPE and the relative MAPE for each of the states and genders as
\begin{align*}
    \text{RMSPE}^{g}_{s}(h) &= \sqrt{\frac{1}{Hp}\sum_{\zeta=h}^{H}\sum_{i=1}^{p} \Biggl[ \frac{\Y^{g}_{T+\zeta, s}(u_i)-\widehat{\Y}^{g}_{T+\zeta, s}(u_i)}{\Y^{g}_{T+\zeta, s}(u_i)}\Biggr]^2\times 100} \\
    \text{MAPE}^{g}_{s}(h) &= \frac{1}{Hp}\sum_{\zeta=h}^{H}\sum_{i=1}^{p}  \Bigg|\frac{\Y^{g}_{T+\zeta, s}(u_i)-\widehat{\Y}^{g}_{T+\zeta, s}(u_i)}{\Y^{g}_{T+\zeta, s}(u_i)}\Bigg| \times 100,
\end{align*}
where $\Y^{g}_{T+\zeta,s}(u_i)$ represents the holdout sample for state $s$ and gender $g$. $\widehat{\Y}^{g}_{T+\zeta,s}(u_i)$ represents the corresponding point forecasts. 

For the considered disaggregation level by state $s$ and population $g$, in \S~\ref{Comp:PF},  we report the average measurement of point forecasts over the whole forecasting horizon $H=10$, leading to a mean  RMSPE and mean  MAPE given by
 \begin{align*}
     \overline{\text{RMSPE}}_s^g &=\frac{1}{H}\sum_{h=1}^{H}\text{RMSPE}^{g}_{s}(h)
\\     
\overline{\text{MAPE}}_s^g &=\frac{1}{H}\sum_{h=1}^{H}\text{MAPE}^{g}_{s}(h).
 \end{align*}

\subsubsection{Point forecast comparison}\label{Comp:PF}

\begin{figure}[!ht]
\centering
\includegraphics[width=8.5cm]{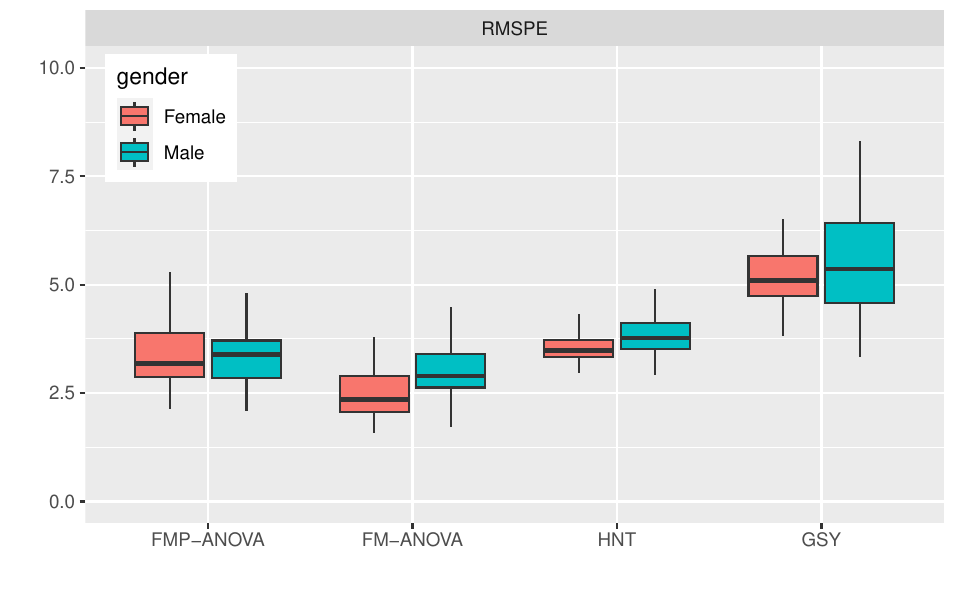}
\qquad
\includegraphics[width=8.5cm]{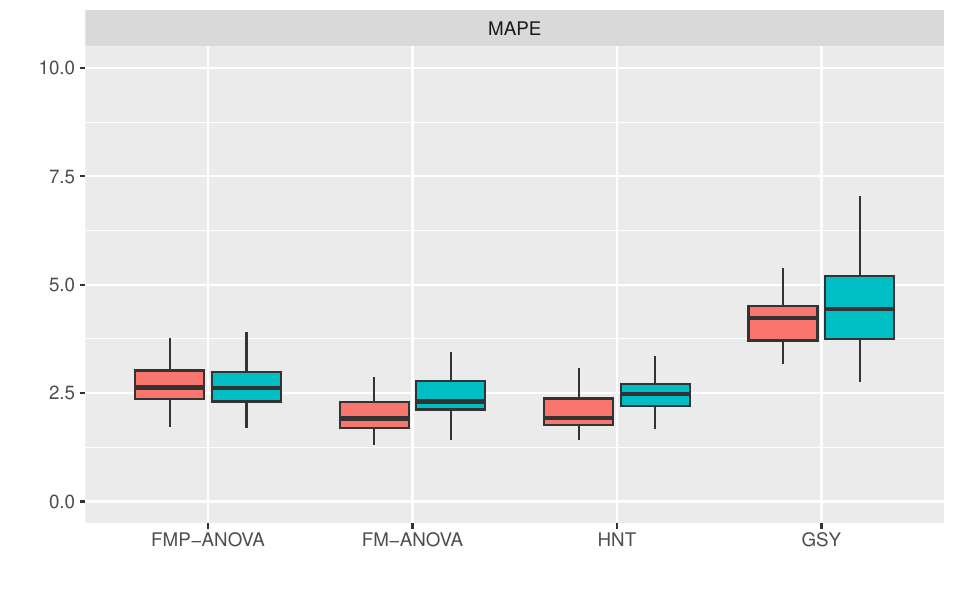}
\\
\includegraphics[width=8.5cm]{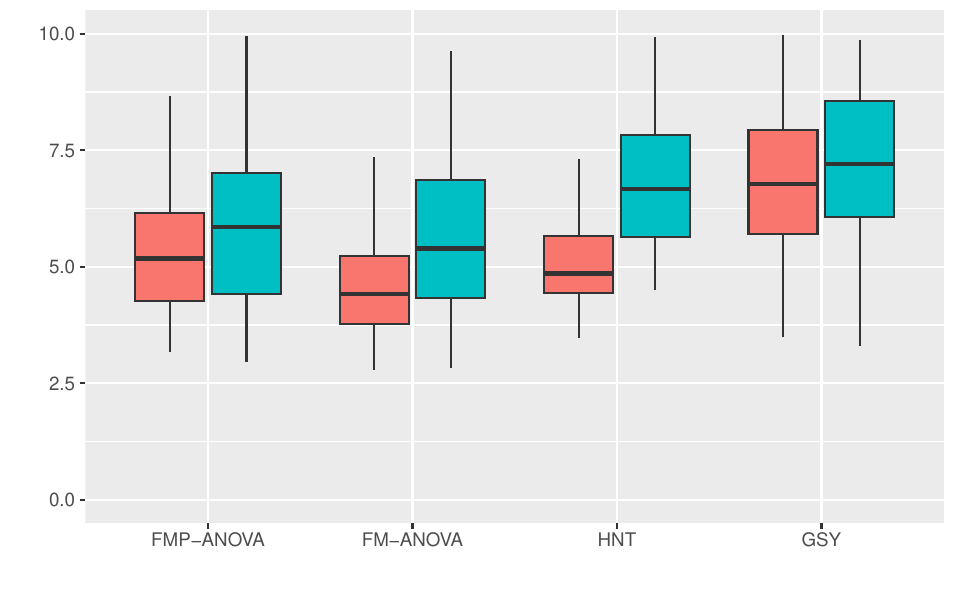}
\qquad
\includegraphics[width=8.5cm]{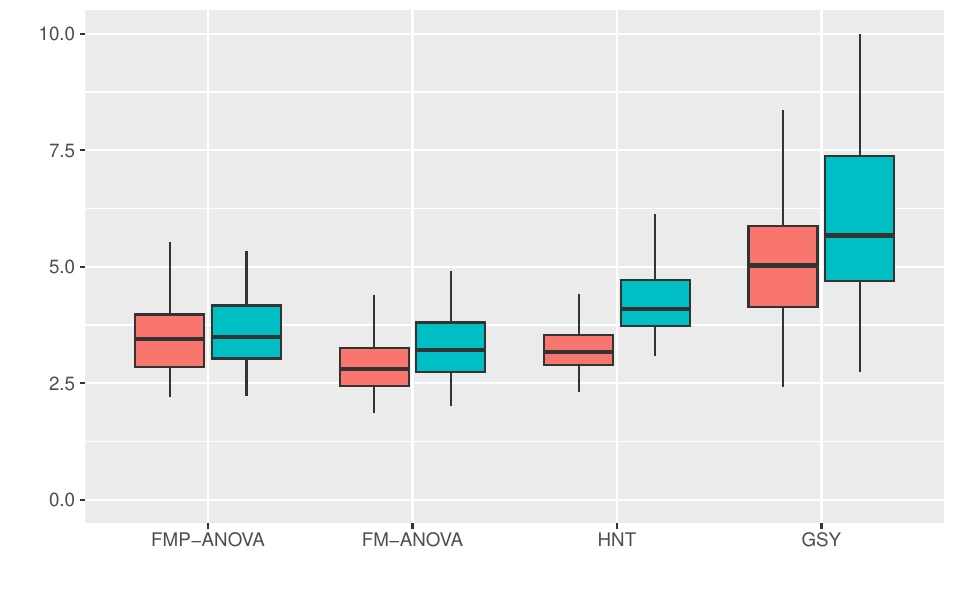}
\\
\includegraphics[width=8.5cm]{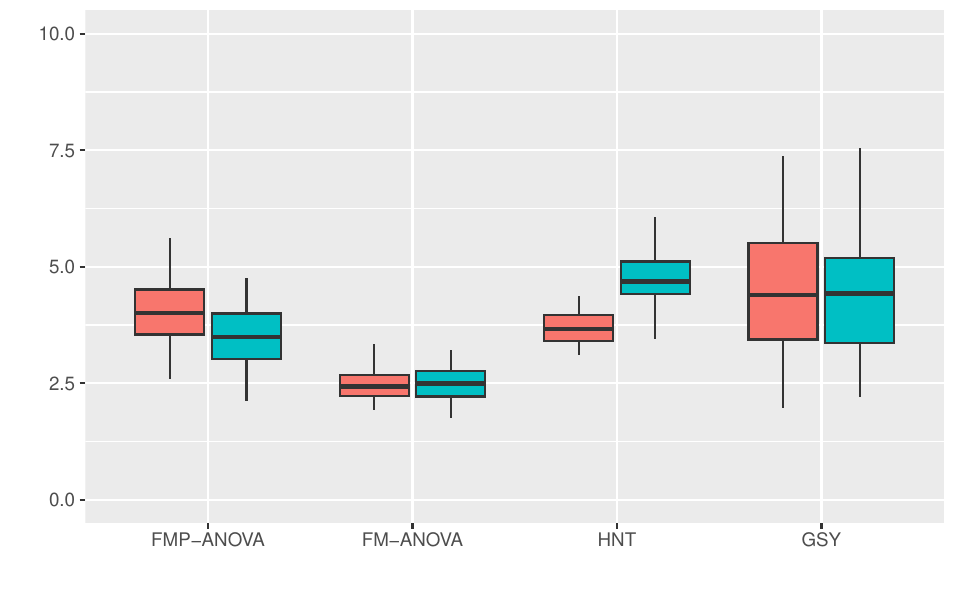}
\qquad
\includegraphics[width=8.5cm]{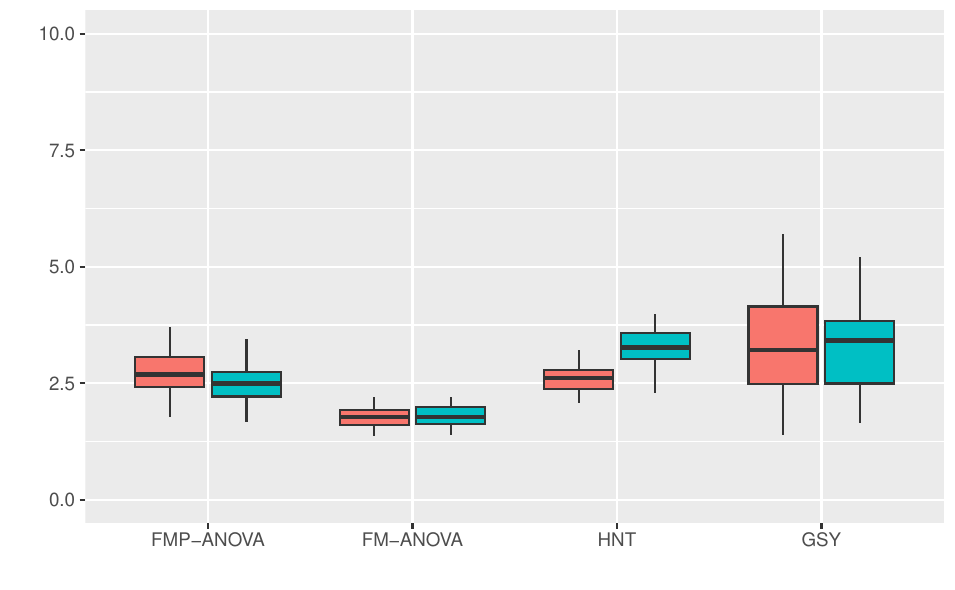}
\caption{ \small{The point forecast error, averaged across states, is calculated based on the estimation of the covariance operator, and the retained components are estimated by the eigenvalue ratio criterion (EVR). The results are presented in three rows: the first row corresponds to the US, the second to France, and the third to Japan. The left column displays the  RMSPE, while the right column displays the  MAPE. HNT refers to the method of \cite{HNT232}, while GSY refers to the method of \cite{GSY19}. }}\label{fig:4}
\end{figure}

We present the results for the point forecasts for the three datasets considered: the US, France, and Japan.  Averaging over the $H=10$ time horizon at each state $s$ and gender $g$, Figure~\ref{fig:4} presents the  $\text{mean}(\text{RMSPE})$ and $\text{mean}(\text{MAPE})$ values using the proposed approaches: FMP-ANOVA (most left), FM-ANOVA (left), the functional factor models from \cite{GSY19} (right), and  \cite{HNT232} (most right). Figure~\ref{fig:4} represents the results for the average obtained by each of the states for the two considered populations, male (in blue) and female (in orange) when forecasting. 

 We can observe that our proposed approach shows better performance for point forecasting in each of the three datasets. In particular, in the first and third rows of Figure~\ref{fig:4}, FM-ANOVA shows a much better point forecast result compared to the FMP-ANOVA, though the FMP-ANOVA attains better results compared to the benchmark methods. The most homogeneous behavior in mortality rates for Japan compared to the US and France is not surprising given Japan's low mortality rates in relation to other G7 nations \citep{tsugane_2020}.  In terms of population, we can see that in the cases of the US and France, male forecasting errors are slightly larger than female forecasting errors.  The individual forecast errors for horizons $h=1,\ldots, H$, obtained from both methods for each state, are available in a developed shiny app \url{https://cristianjv.shinyapps.io/HDFTSForecasting/}.

\subsection{Interval forecast evaluation} \label{IF_comp}

To evaluate pointwise interval forecast accuracy, we consider the coverage probability difference (CPD) between the nominal and empirical coverage probabilities. The empirical coverage probability is defined as follows 
\begin{align*}
    \text{Empirical coverage}^{g}_{s}=1-\frac{1}{H p} 
    \sum_{\zeta=h}^{H} \sum_{i=1}^p \Bigl[ &\mathbbm{1} \Bigl\{ \Y^{g}_{T+\zeta|T,s}(u_i) > \widehat{\Y}^{g,\text{ub}}_{T+\zeta|T, s}(u_i) \Bigr\} + \\ 
    & \mathbbm{1}\Bigl\{\Y^{g}_{T+\zeta|T, s}(u_i)< \widehat{\Y}^{g,\text{lb}}_{T+\zeta|T, s}(u_i)\Bigr\} \Bigr],
\end{align*}
where $H$ denotes the number of curves in the forecasting period, $p$ denotes the number of discretized points for the age, $\widehat{\Y}^{g, \text{ub}}_{T+\zeta|T, s}$ and $\widehat{\Y}^{g, \text{lb}}_{T+\zeta|T, s}$ denote the upper and lower bounds of the corresponding forecasted interval, and $\mathbbm{1}\{\cdot\}$ the binary indicator function. Pointwise CPD is defined as
\begin{equation*}    \text{CPD}^{g}_{s}=\Bigg|\text{Empirical coverage}^{g}_{s}-\text{Nominal coverage} \Bigg|.
\end{equation*}
The lower the CPD$^{g}_{s}$ value, the better the forecasting method's performance. 

Additionally, we utilize the interval score of \cite{Gneiting2007} \citep[see also][]{GK14}. For each year in the forecasting period, the $h$-step-ahead prediction intervals were calculated at the $100(1-\alpha)\%$ nominal coverage probability. We consider the common case of the symmetric $100(1-\alpha)\%$ prediction interval, with lower and upper bounds that are predictive quantiles at $\alpha/2$ and $1-\alpha/2$, denoted by $ \widehat{\Y}^{g, \text{lb}}_{T+\zeta|T, s}(u_i)$ and $\widehat{\Y}^{g, \text{ub}}_{T+\zeta|T, s}(u_i)$. The scoring rule for the interval forecast at discretized point $u_i$ is
\begin{align*}
    S^{g}_{\alpha,\zeta,s}&\left[\widehat{\Y}^{g, \text{lb}}_{T+\zeta|T,s}(u_i),\widehat{\Y}^{g, \text{ub}}_{T+\zeta|T,s}(u_i),\Y^{g}_{T+\zeta|T,s}(u_i)  \right] = \left[\widehat{\Y}^{g, \text{ub}}_{T+\zeta|T,s}(u_i)-\widehat{\Y}^{g, \text{lb}}_{T+\zeta|T,s}(u_i) \right]\\
    & +\frac{2}{\alpha}\left[\widehat{\Y}^{g, \text{lb}}_{T+\zeta|T,s}(u_i)-\Y^{g}_{T+\zeta|T,s}(u_i) \right] \mathbbm{1}\Bigl \{ \Y^{g}_{T+\zeta|T,s}(u_i) < \widehat{\Y}^{g, \text{lb}}_{T+\zeta|T,s}(u_i)\Bigr \}\\
    &+\frac{2}{\alpha}\left[\Y^{g}_{T+\zeta|T,s}(u_i)-\widehat{\Y}^{g, \text{ub}}_{T+\zeta|T,s}(u_i) \right] \mathbbm{1}\Bigl \{ \widehat{\Y}^{g, \text{ub}}_{T+\zeta|T,s} (u_i)>\Y^{g}_{T+\zeta|T,s}(u_i)  \Bigr \},
\end{align*}
where $\mathbbm{1}\{\cdot\}$ represents the binary indicator function, and $\alpha$ denotes a level of significance. Finally, we compute the mean interval score for the total of $T$ series as
\begin{equation*}
    \overline{S}^{g}_{\alpha,s}=\frac{1}{H p}\sum_{\zeta=h}^{H}\sum_{i=1}^{p}S_{\alpha,\zeta,s}^g\left[\widehat{\Y}^{ \text{g,\text{lb}}}_{T+\zeta|T,s}(u_i),\widehat{\Y}^{g, \text{ub}}_{T+\zeta|T,s}(u_i),\Y^{g}_{T+\zeta|T,s}(u_i) \right].
\end{equation*}
The optimal interval score is achieved when $\Y^{g}_{T+\zeta|T,s}(u_i)$ lies between $\widehat{\Y}^{g, \text{lb}}_{T+\zeta|T,s}(u_i)$  and $\widehat{\Y}^{g, \text{ub}}_{T+\zeta|T,s}(u_i)$,  with the distance between the upper bound and the lower bound being minimal.

\subsubsection{Interval forecast comparison}\label{Comp:IF_FMP}

\begin{figure}[htbp]
\centering
\includegraphics[width=8.5cm]{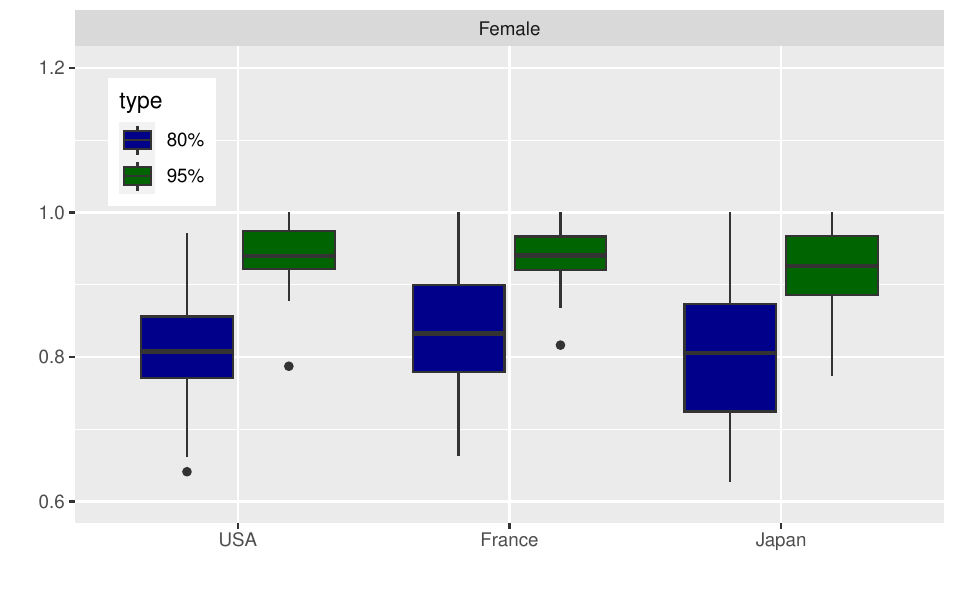}
\qquad
\includegraphics[width=8.5cm]{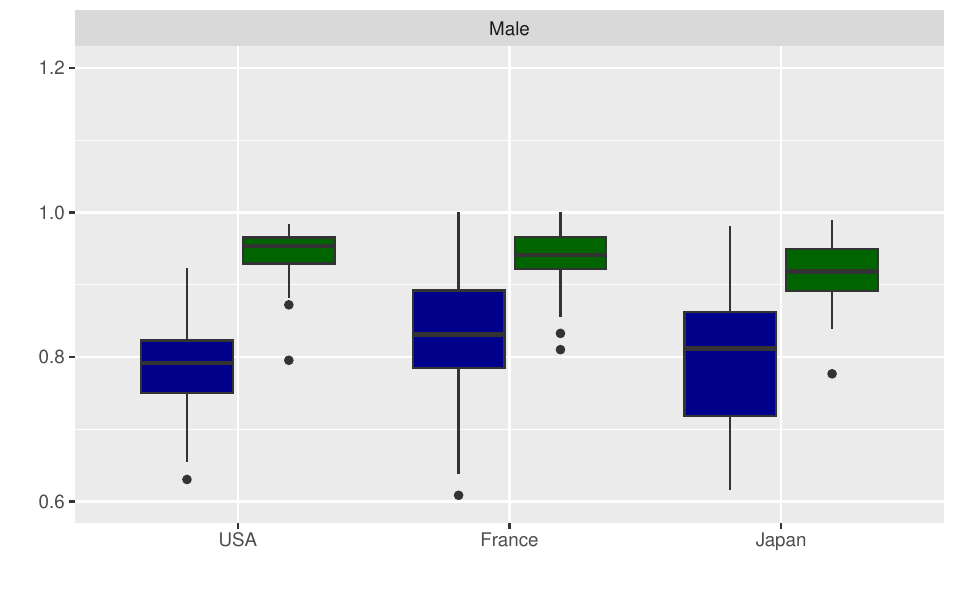}
\\
\includegraphics[width=8.5cm]{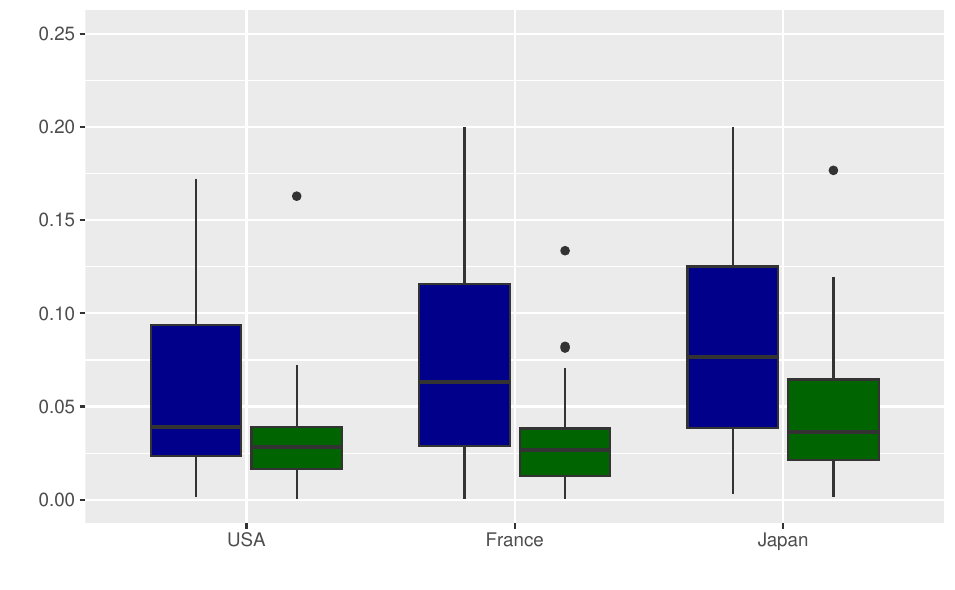}
\qquad
\includegraphics[width=8.5cm]{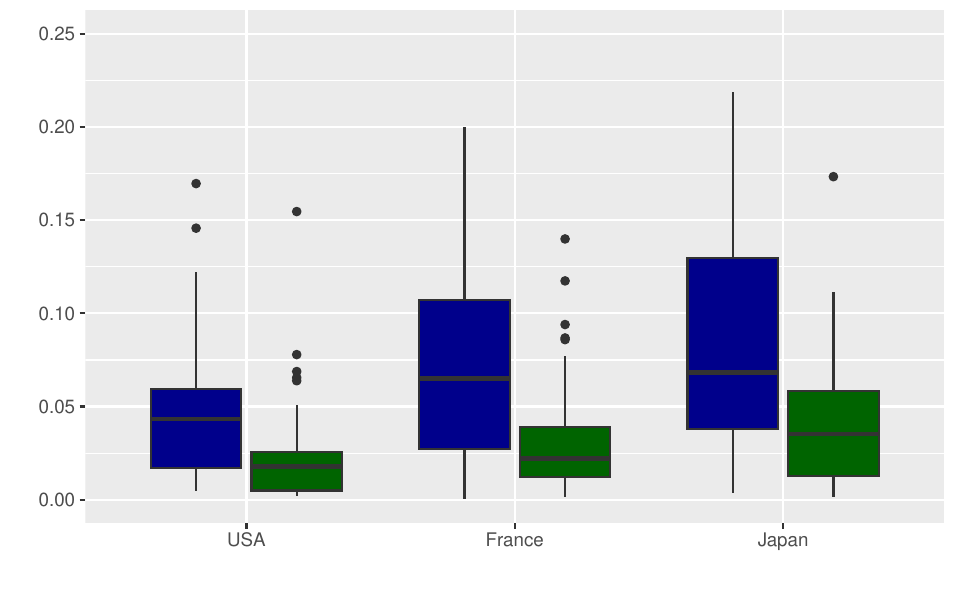}
\\
\includegraphics[width=8.5cm]{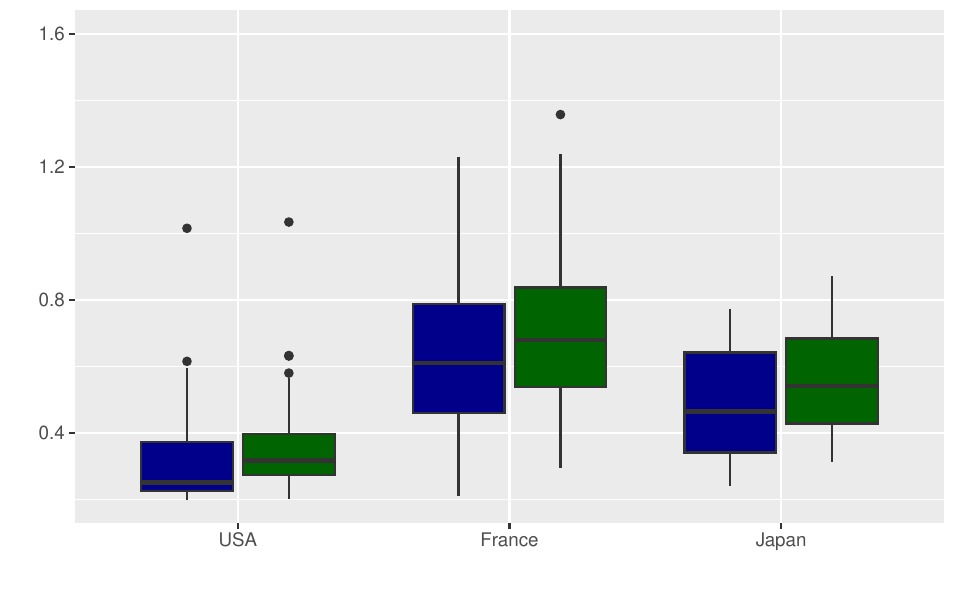}
\qquad
\includegraphics[width=8.5cm]{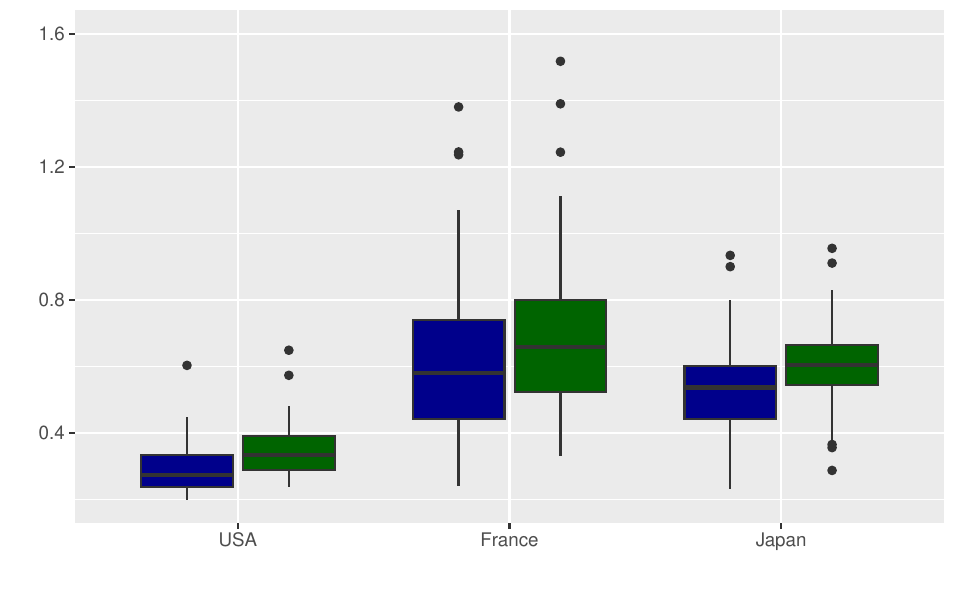}
\caption{\small{The plot displays the results of interval prediction performance across states for all three countries using the FMP-ANOVA decomposition. The top row represents the pointwise empirical coverage probability. The middle row shows the coverage probability difference (CPD). The bottom row presents the mean interval score. The left panel corresponds to the female population, while the right panel corresponds to the male population. Two nominal coverage probabilities are considered: $80\%$ and $95\%$. Each plot includes data for the United States (leftmost), France (center), and Japan (rightmost).}}\label{fig:5}
\end{figure}

\begin{figure}[!htbp]
\centering
\includegraphics[width=8.5cm]{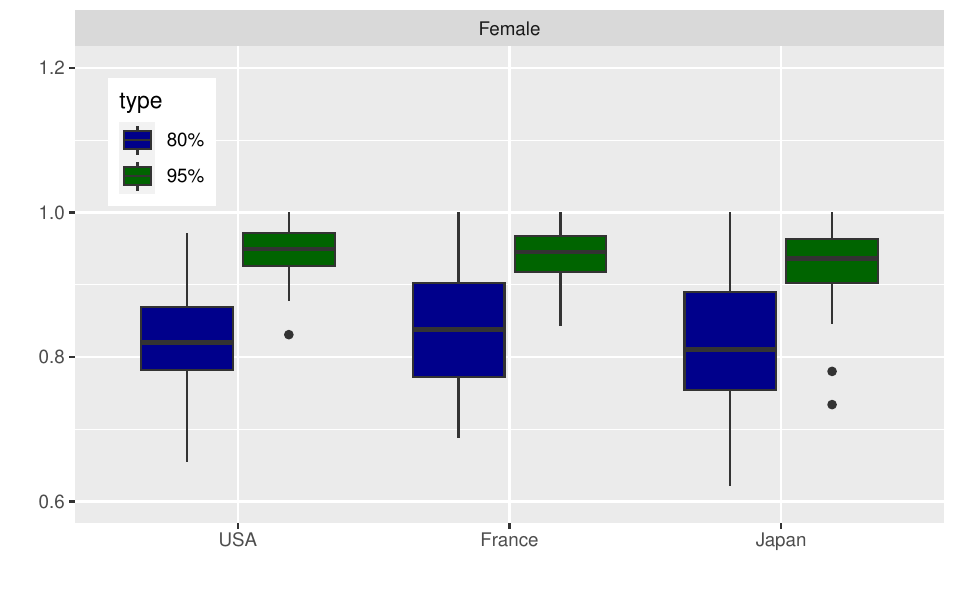}
\qquad
\includegraphics[width=8.5cm]{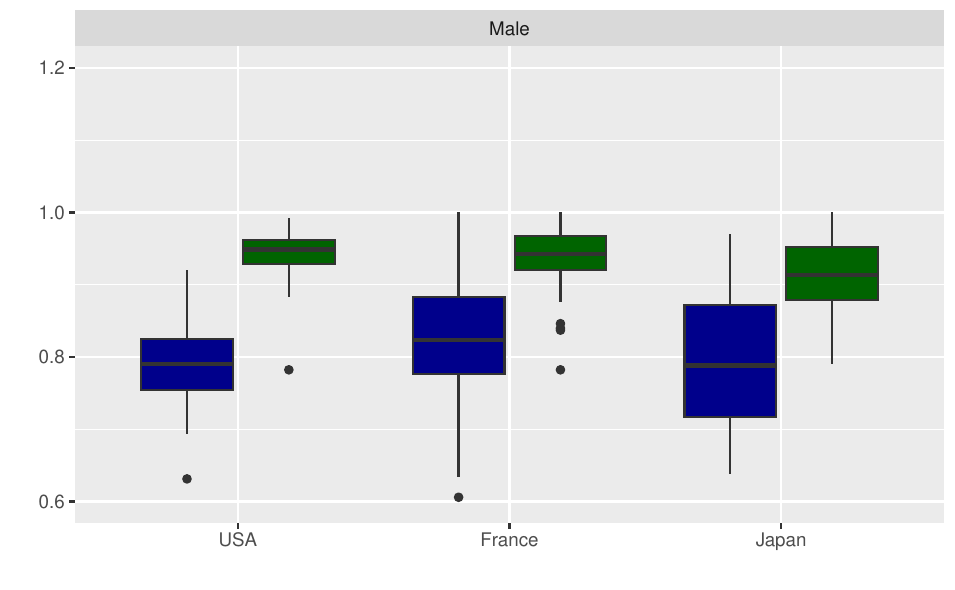}
\\
\includegraphics[width=8.5cm]{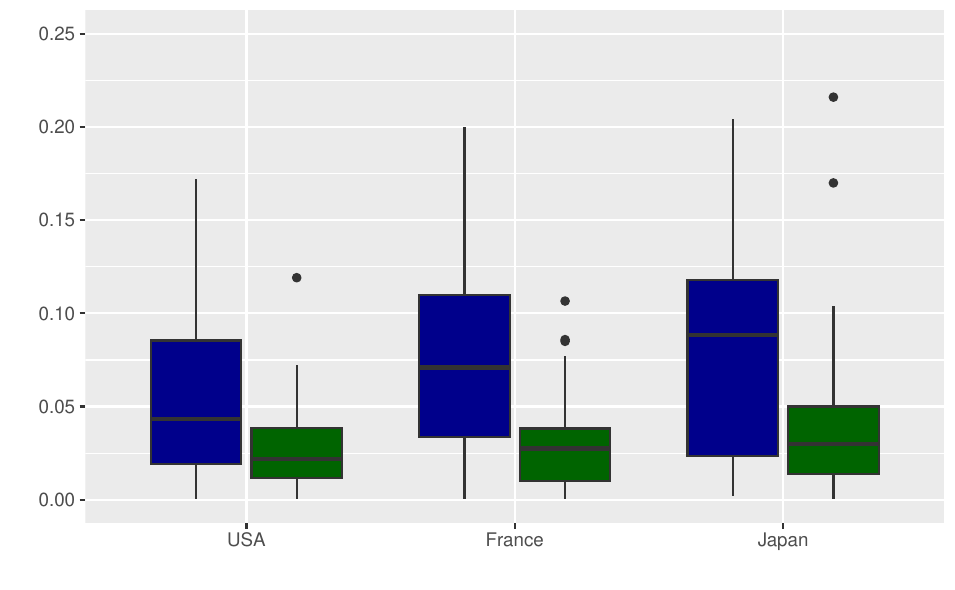}
\qquad
\includegraphics[width=8.5cm]{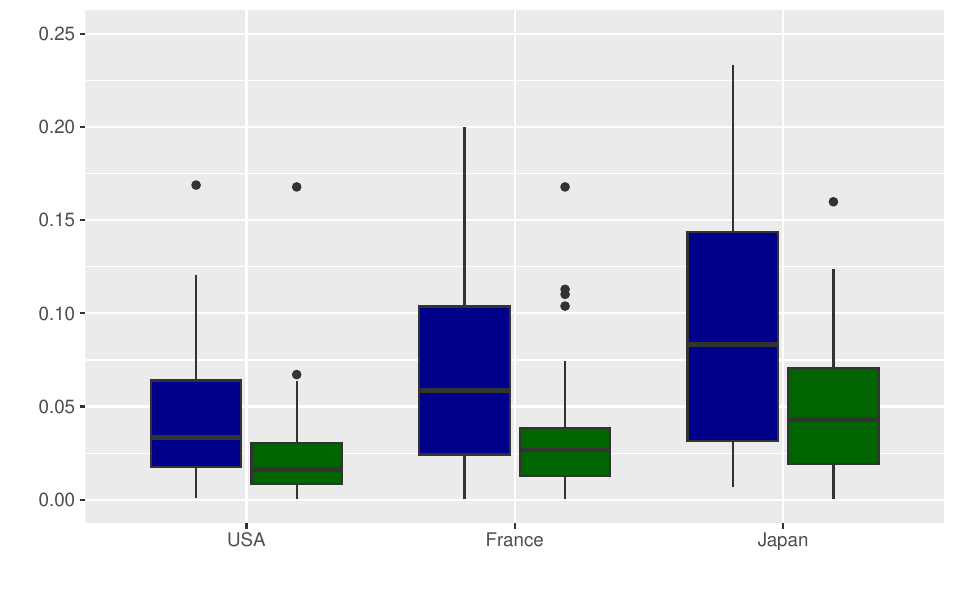}
\\
\includegraphics[width=8.5cm]{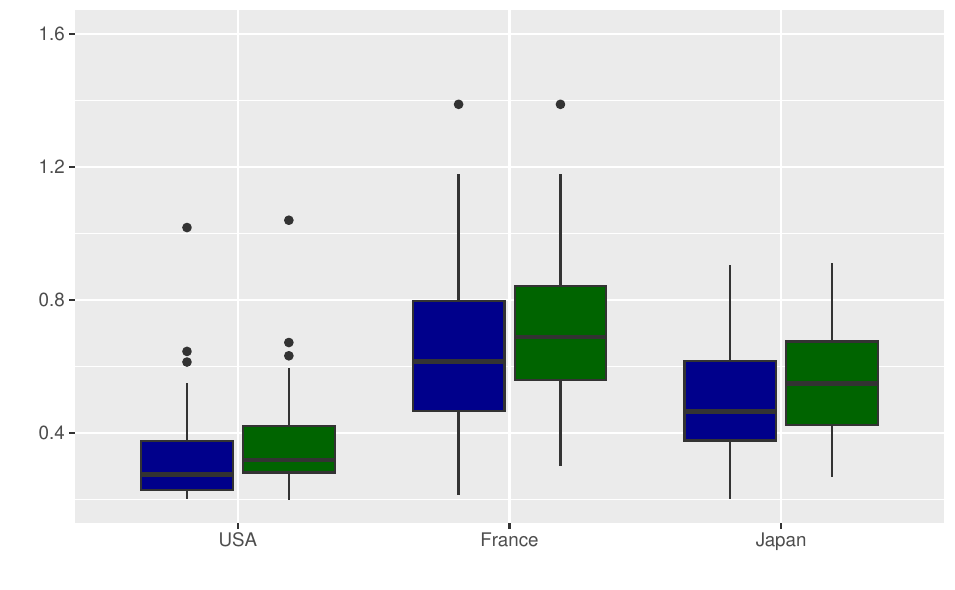}
\qquad
\includegraphics[width=8.5cm]{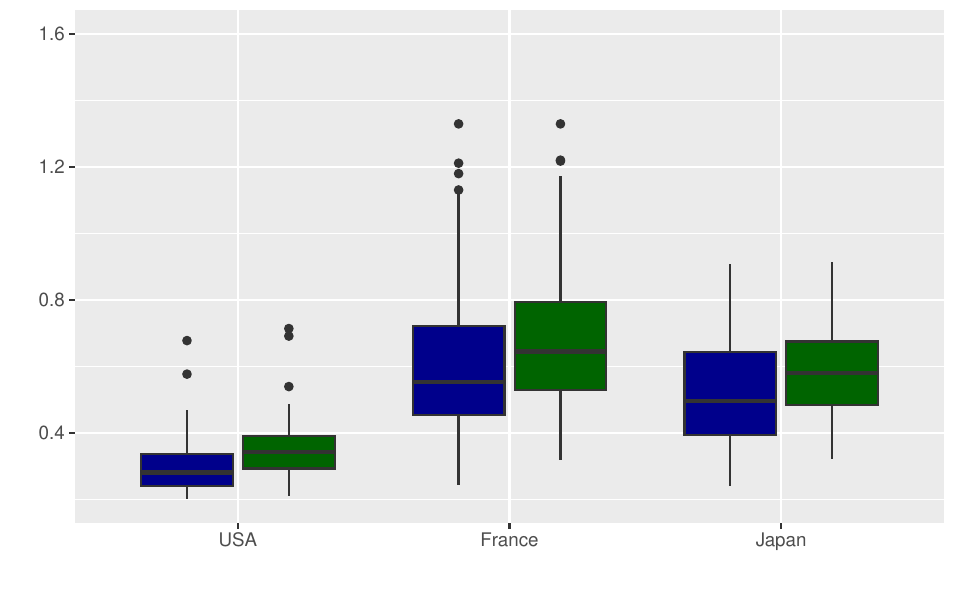}
\caption{\small{The plot displays the results of interval prediction performance across states for all three countries using the FM-ANOVA decomposition. The top row represents the pointwise empirical coverage probability. The middle row shows the coverage probability difference (CPD). The bottom row presents the mean interval score. The left panel corresponds to the female population, while the right panel corresponds to the male population. Two nominal coverage probabilities are considered: $80\%$ and $95\%$. Each plot includes data for the United States (leftmost), France (center), and Japan (rightmost).}}\label{fig:6}
\end{figure}

We present the interval forecast results for the three datasets:  the US, France, and Japan. Interval forecasts are assessed when constructed using both the proposed FMP-ANOVA (Figure~\ref{fig:5}) and FM-ANOVA (Figure~\ref{fig:6}) approaches. The averages correspond to each of the states for all data examples. We present the averages across the forecasting horizon of $H=10$ years ahead for two different interval forecasting accuracies: empirical coverage probability (top panel), coverage probability difference (CPD) (middle panel); and mean interval score (bottom panel). 

 In the first row of Figures~\ref{fig:5} and~\ref{fig:6}, we present the pointwise coverage probability for both male and female populations. For each of the countries, we consider two nominal coverages: $80\%$  and $95\%$. For the FMP-ANOVA approach in Figure~\ref{fig:5},  by examining the median level of the averages for the pointwise coverage probability, we can see that all three nations and both populations are very close to the nominal levels. However, the FMP-ANOVA approach performs better for the French dataset towards the $95\%$ nominal level than the $80\%$ nominal level for both males and females. The FMP-ANOVA approach outperforms the case of the US compared to the other two countries in population and nominal levels. The Japanese dataset achieves a  level similar to the $80\%$, although the median level in the $95\%$ example seems to be lower. 

In contrast, when using the FM-ANOVA in Figure~\ref{fig:6}, we can observe that for France, the $80\%$ nominal level at the median level of the empirical coverage probability is upper skewed, while for the other two countries, it remains very similar to the nominal level, performing better. Similar to the results from the FMP-ANOVA, we can observe that the forecast intervals with the $95\%$ nominal level perform much better.

In the same line as with the pointwise coverage probability, and to make the interpretation of the results easier, in the second row of Figures~\ref{fig:5} and~\ref{fig:6}, we present the CPD for both populations as well as both nominal levels. In general, we can observe that for the three countries, the $95\%$ nominal level achieves the lowest range of values of the CPD. Regarding the $80\%$ nominal level, the FMP-ANOVA produced superior results in the case of the US over the case of France. In contrast, in the case of Japan, both approaches outperformed France. In Figure~\ref{fig:6} with the $95\%$ nominal level, FM-ANOVA yields the narrowest range of values for the cases of the US and France, which outperform the Japanese case in terms of CPD. The third row of Figures~\ref{fig:5} and~\ref{fig:6} presents the results for the mean interval score for both populations. In general, according to both methodologies, the interval score shows better performance in the case of  France than in the other two countries; nonetheless, the case of the US has the best overall average interval score performance. 

\section{Conclusion}\label{sec:conclusion}

We have proposed an innovative FTS forecasting methodology based on the two-way functional median polish. The proposed strategy is useful for FTS models with complex structures, particularly those involving states and various populations. This method of forecasting grouped FTS is derived by combining an estimated two-way functional ANOVA model with the FPCA framework. Using age-specific mortality rates at the national and sub-national levels in the US, France, and Japan, we compare the averages across the $10$ point forecast accuracies for the proposed method based on functional-ANOVA decomposition with several benchmark methods like the factor models of \cite{GSY19} and \cite{HNT232} and independent FTS (see Appendix~\ref{A2}). We can see that the two-way functional ANOVA decomposition strategy beats benchmark methods.

There are several ways in which the present methodology can be further extended, and we briefly mention four.
\begin{asparaenum}
\item[1)] First, one restriction of the proposed method is the possibility of outliers, which may significantly impact the modeling and forecasting of principal component scores. Using a robust functional principal component approach \citep[see, e.g.,][]{Bali2011} or other robust time series methods \citep[see, e.g.,][]{Gelper2010} are possible approaches for addressing this issue.
\item[2)] Second, other levels of disaggregation may be included in the suggested approach with the availability of appropriate data. Cause-of-death, as mentioned in \cite{Gaille2015} and socioeconomic status \citep{Singh2013} are examples of such levels. In the same spirit, we may incorporate different mortality data measures, such as the age distribution of death counts as in \cite{shang_haberman_2020} and \cite{Shang2022}. 
\item[3)] Third, the proposed methodology may be employed in other application areas, such as university performance completion rates, which can be disaggregated by age, gender, faculty, local or international status, and other criteria. Such disaggregation levels enable us to employ joint forecast approaches that use constrained estimates of age-specific completion rates to quantify the effect of various factors that may explain completion behavior.
\item[4)] The functional residual component, which exhibits time-varying characteristics and is derived from estimating the two-way functional ANOVA model, can be regarded as a functional time series (FTS) on its own. Consequently, employing functional factor models, such as the one proposed by \cite{HNT232}, is a viable approach to enhance the forecasting accuracy of this component.

\end{asparaenum}

\section*{Supplementary Materials}

\textbf{Code for FTS forecasting based on FMP-ANOVA and FM-ANOVA.} The R code to produce point and interval forecasts from the two approaches described in the paper. The R codes are available at the following repository: \url{https://github.com/cfjimenezv07/Forecasting_HDFTS/tree/main/Rcodes_paper}\\
\textbf{Code for shiny application.} The R code to produce a shiny user interface for plotting every series and the results for point and interval forecasts for the three considered mortality databases. The R codes are available at the following repository: \url{https://github.com/cfjimenezv07/Forecasting_HDFTS/tree/main/Shiny_app}

\section*{Acknowledgement}

The first author acknowledges the financial support of the King Abdullah University of Science and Technology (KAUST). The last author acknowledges the funding of an Australian Research Council Discovery Project DP230102250 titled ``Feature learning for high-dimensional functional time series" and Macquarie University DataX consilience center. The authors are grateful for the comments from the participants at the Australian National University, University of Auckland, Australian Government Actuary, the 6\textsuperscript{th} International Conference on Econometrics and Statistics (EcoSta 2023), Joint Statistical Meeting, and Australian Statistical Conference in 2023.

\bibliographystyle{chicago}
\bibliography{paper4}

\newpage
\appendix

\section{Appendix}

\subsection{Multivariate time series forecasting approach.}\label{A1}

\begin{figure}[!htb]
\centering
{\includegraphics[width=7.1cm]{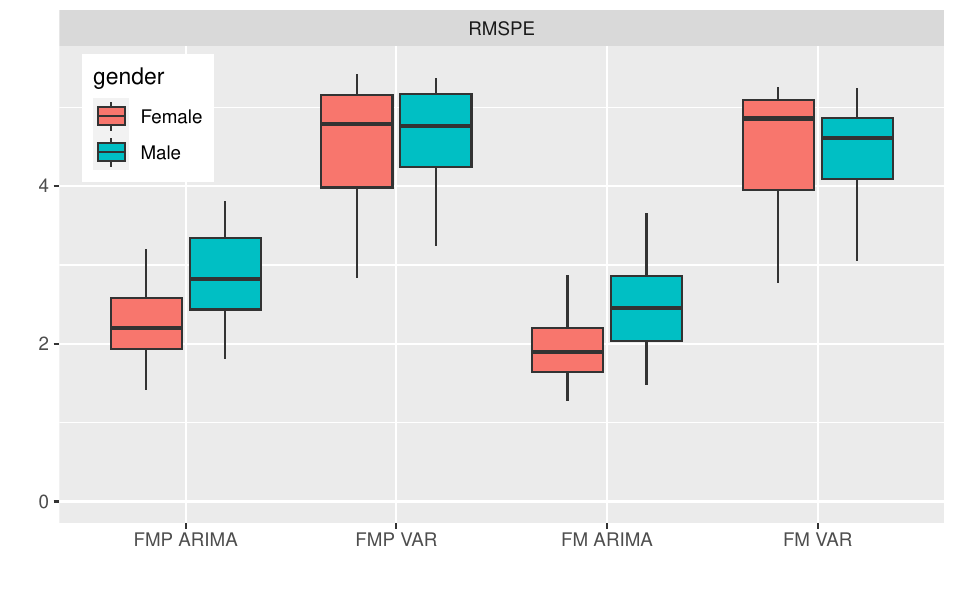}}
\qquad
{\includegraphics[width=7.1cm]{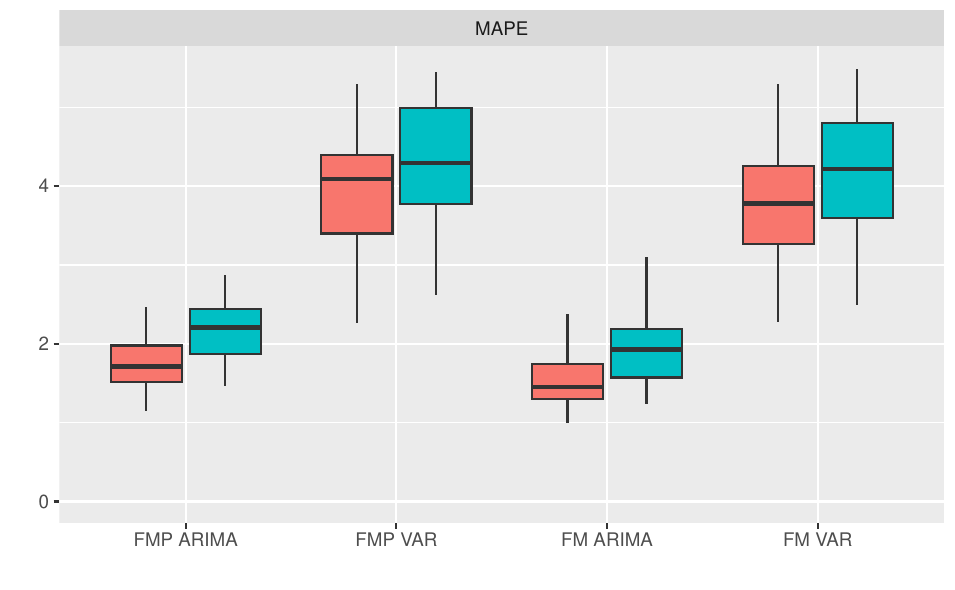}}
\\
{\includegraphics[width=7.1cm]{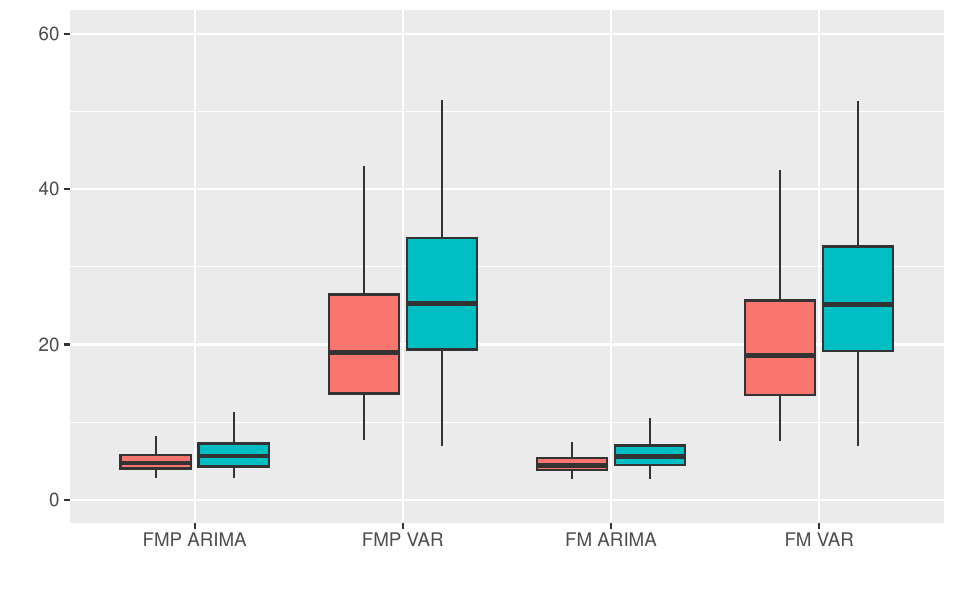}}
\qquad
{\includegraphics[width=7.1cm]{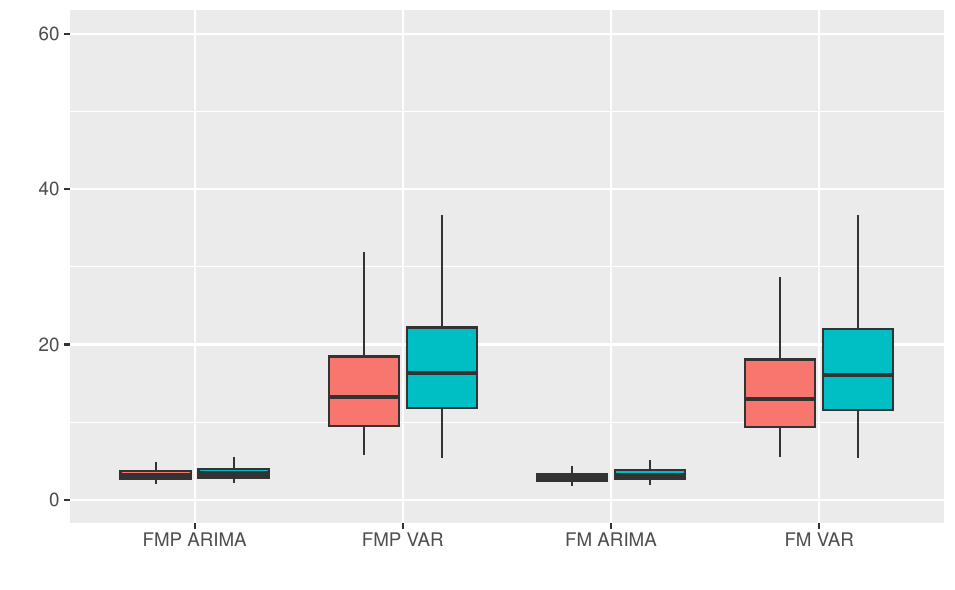}}
\\
{\includegraphics[width=7.1cm]{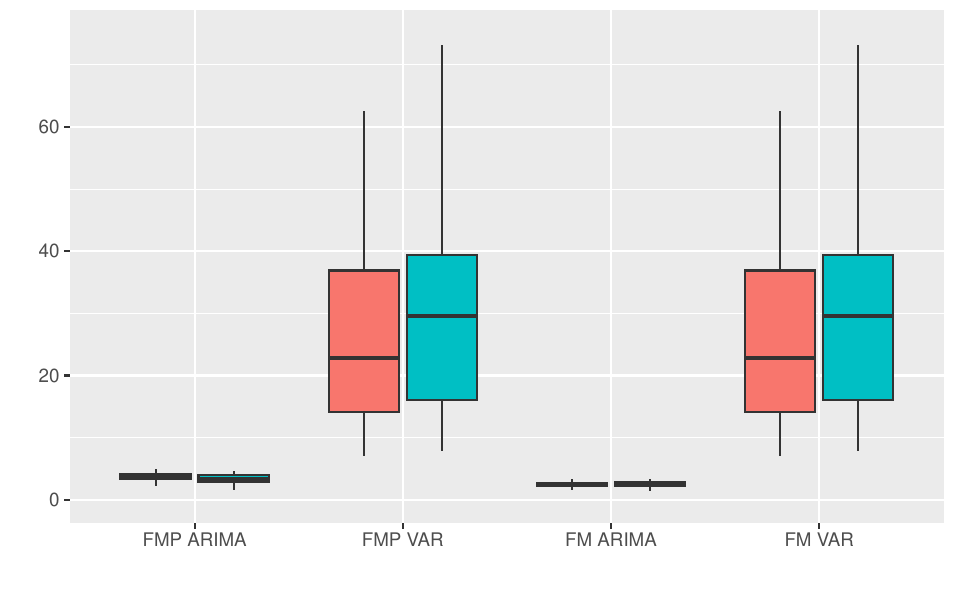}}
\qquad
{\includegraphics[width=7.1cm]{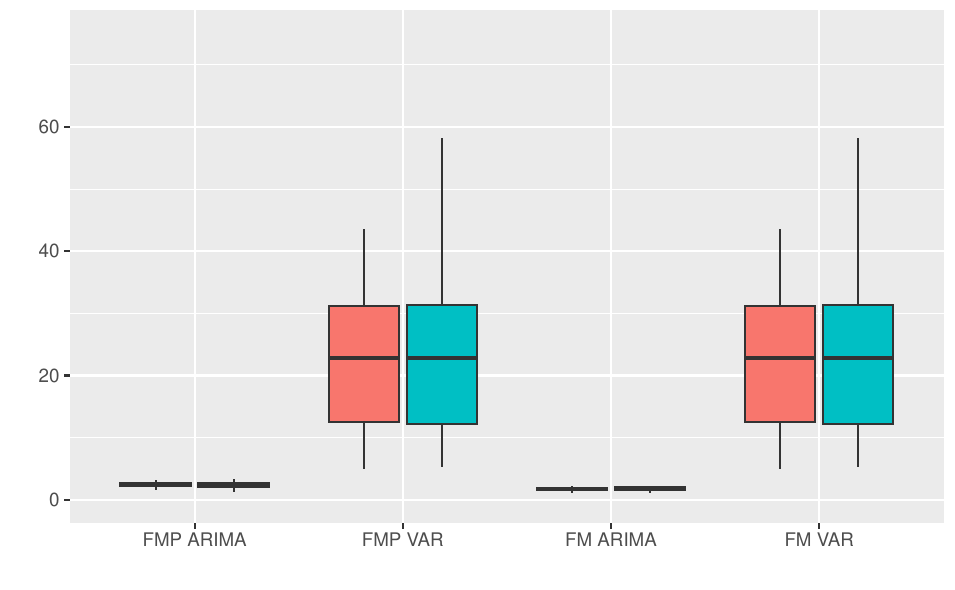}}
\caption{\small Comparison between the univariate ARIMA and multivariate VAR time series forecasting methods when $K=6$.  The results are presented in three rows: the first row corresponds to the US, the second row to France, and the third row to Japan. The left column displays the relative RMSPE, while the right column displays the relative MAPE.}\label{fig:6_uni_vs_mult}
\end{figure}

We conducted an additional analysis by fixing the number of retained components at $K=6$, as it has been shown that this number of components is enough \citep{H2013}. In this scenario, we explored the application of a multivariate time series forecasting method, specifically VAR, to conduct the forecasting. To facilitate comparison, Figure~\ref{fig:6_uni_vs_mult} presents a comparison between the univariate ARIMA and multivariate VAR time series forecasting methods. Notably, the results from the univariate time series method (ARIMA) demonstrate notably superior forecasting performance.

\clearpage
\subsection{Independent populations}\label{A2}

\begin{figure}[!ht]
\centering
\includegraphics[width=8.5cm]{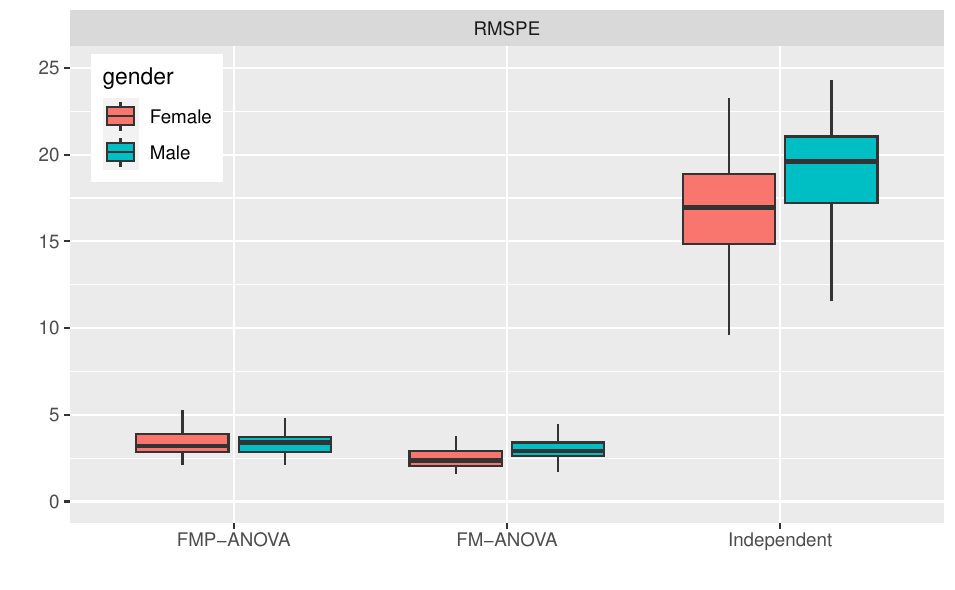}
\quad
\includegraphics[width=8.5cm]{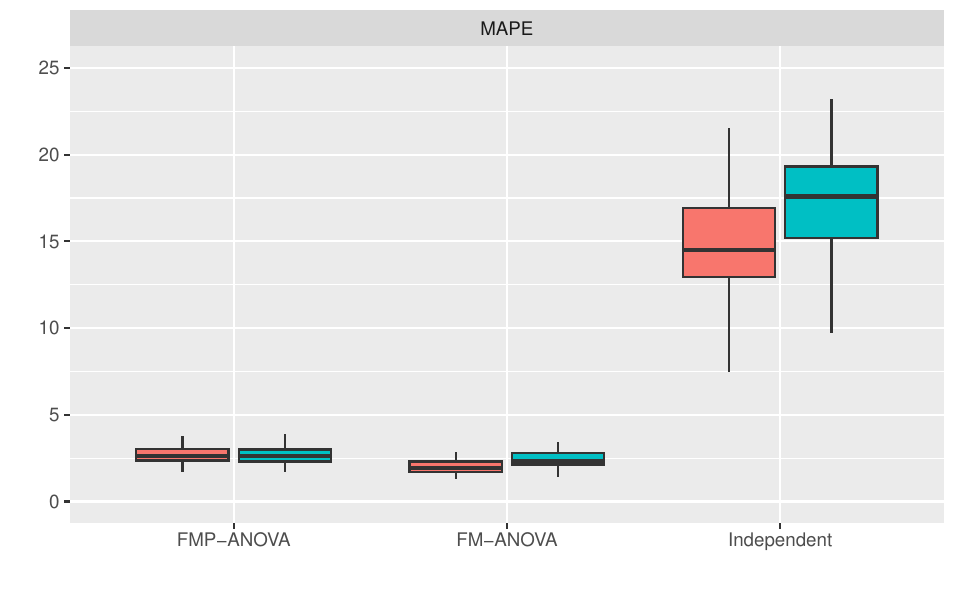}
\\
\includegraphics[width=8.5cm]{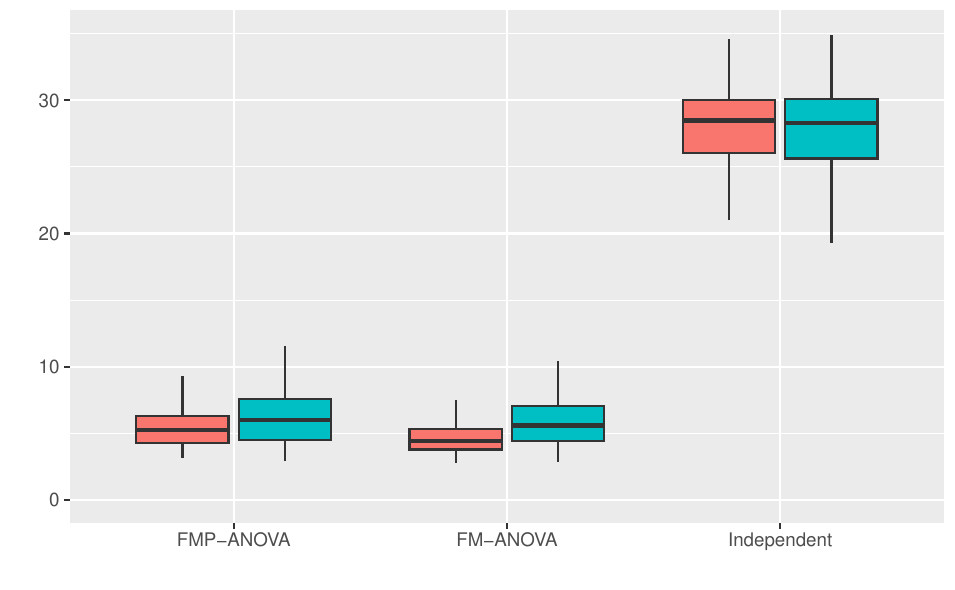}
\quad
\includegraphics[width=8.5cm]{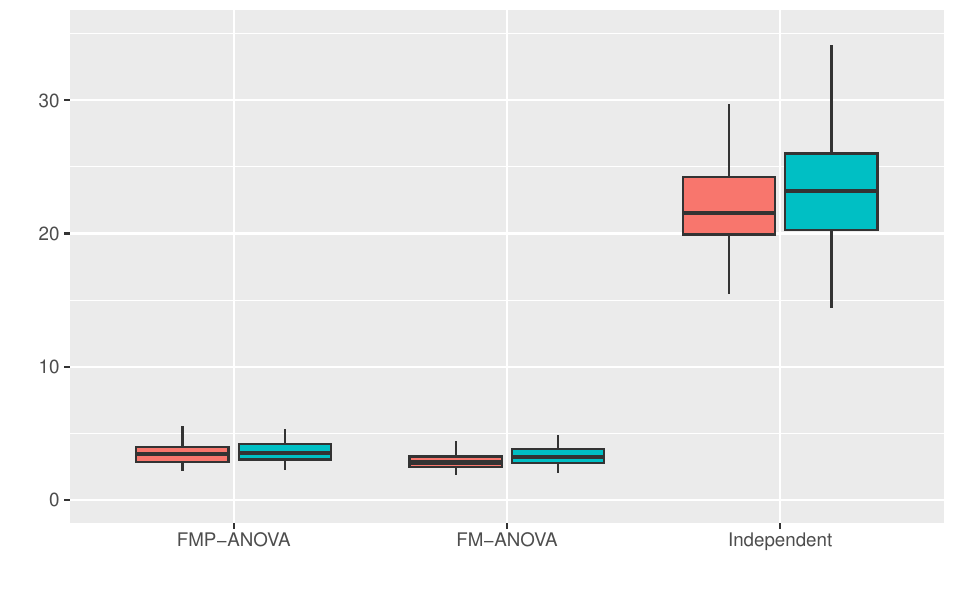}
\\
\includegraphics[width=8.5cm]{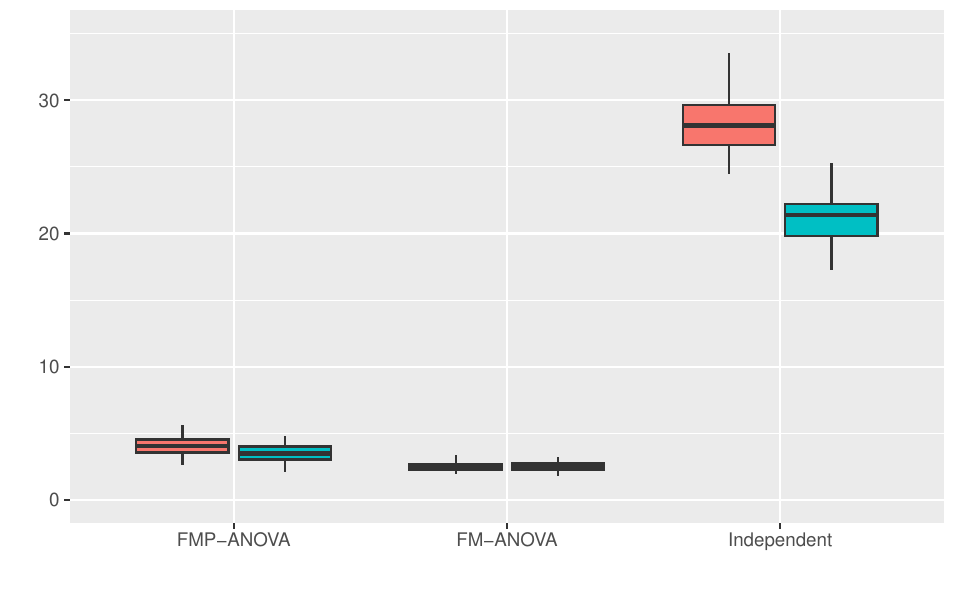}
\quad
\includegraphics[width=8.5cm]{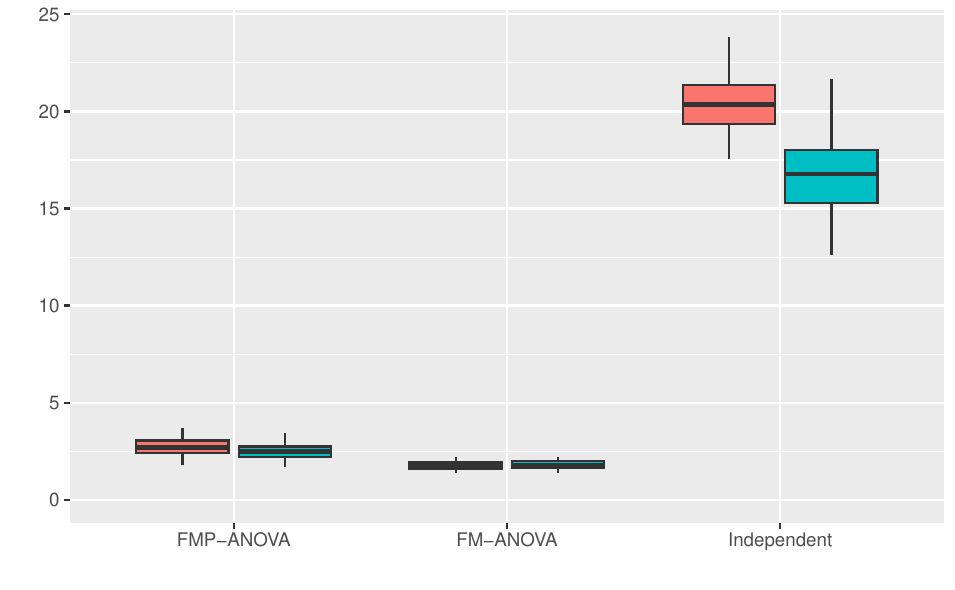}
\caption{\small{The US average forecast error per state, the French average forecast error per department, and the Japanese average forecast error per prefecture. The relative RMSPE is shown to the left, while the relative MAPE is to the right.}}\label{fig:9}
\end{figure}

In order to establish a real enhancement of the forecasting approach based on the estimation of a two-way functional ANOVA model, in this Appendix, we present the point forecast results of the two proposed estimating methods: FMP-ANOVA and FM-ANOVA in comparison to a naive approach in which all populations are considered to be independent. In Figure~\ref{fig:9}, we can observe the point forecast errors are larger for all states in the three datasets for the independent forecasting method.

\end{document}